\documentclass[a4paper,fleqn]{cas-dc}

\usepackage{natbib}
\usepackage{svg}
\usepackage{subfig}
\def\tsc#1{\csdef{#1}{\textsc{\lowercase{#1}}\xspace}}
\tsc{WGM}
\tsc{QE}
\tsc{EP}
\tsc{PMS}
\tsc{BEC}
\tsc{DE}

\begin{document}
\let\WriteBookmarks\relax
\def\floatpagepagefraction{1}
\def\textpagefraction{.001}

\shorttitle{WhatsApp Tiplines in the 2021 Indian Assembly Elections}

\shortauthors{Shahi \& Hale}

\title [mode = title]{WhatsApp Tiplines and Multilingual Claims in the 2021 Indian Assembly Elections}                      




\hyphenation{Whats-App}

%
\author[1]{Gautam Kishore Shahi}[orcid=0000-0001-6168-0132]

\cormark[1]
\ead{gautam.shahi@uni-due.de}



\affiliation[1]{organization={University of Duisburg-Essen},
    country={Germany}}

\author[2]{Scott A. Hale}[orcid=0000-0002-6894-4951]

\affiliation[2]{organization={Meedan \& University of Oxford},
    country={United Kingdom}}

\cortext[cor1]{Corresponding author}

\captionsetup[subfigure]{skip=2pt}  


\begin{abstract}
WhatsApp 
tiplines, first launched in 2019 to combat misinformation, enable users to interact with fact-checkers to verify misleading content. This study analyzes 580 unique claims (tips) from 451 users, covering both high-resource languages (English, Hindi) and a low-resource language (Telugu) during the 2021 Indian assembly elections using a mixed-method approach. We categorize the claims into three categories, election, COVID-19, and others, and observe variations across languages. We compare content similarity through frequent word analysis and clustering of neural sentence embeddings. We also investigate user overlap across languages and fact-checking organizations. We measure the average time required to debunk claims and inform tipline users. Results reveal similarities in claims across languages, with some users submitting tips in multiple languages to the same fact-checkers. Fact-checkers generally require a couple of days to debunk a new claim and share the results with users.  Notably, no user submits claims to multiple fact-checking organizations, indicating that each organization maintains a unique audience. We provide practical recommendations for using tiplines during elections with ethical consideration of users' information. 
\end{abstract}


\begin{highlights}
\item Our study focuses on fact-checking on an encrypted platform (WhatsApp) during the 2021 Indian Assembly Elections.
\item We provide analysis of users' claims and fact-checkers' responses in WhatsApp tiplines during the 2021 Indian Assembly Elections.
\item We analyze tipline claims during the election in three languages (Hindu, English \& Telugu), one of which is a low-resource language (Telugu).
\end{highlights}


\maketitle

\section{Introduction}\label{sec1}

High Internet penetration and mobile phone adoption in India have rapidly increased the number of Internet users and the amount of user-generated content on social media platforms \citep{shahi2023exploratory}. Social media sites have become the platforms of choice for people worldwide to create and share information and misinformation during important local and global events \citep{aimeur2023fake}. However, platforms such as Twitter (Now \emph{X}) is primarily designed with a focus on Western users \citep{shahi2022mitigating}, often lacking robust support for many underrepresented languages. As a result, misinformation studies have predominantly focused on Western languages such as English and German, with only a few addressing Asian languages, e.g., Hindi, and none examining regional languages (e.g., Telugu). 


As an encrypted platform, WhatsApp is more difficult to examine and often understudied, possibly due to its end-to-end encryption. 
One way fact-checkers have worked on WhatsApp is via misinformation tiplines. These are run by fact-checking organizations and allow users to send claims (or tips) to fact-checking organizations for debunking. 
Tiplines are particularly important where WhatsApp has a large market penetration and a massive user base. For example, in India, WhatsApp has 97.1\% market penetration with over 400 million users \citep{Statista}. 
While many fact-checking organizations are on WhatsApp,\footnote{\url{https://faq.whatsapp.com/5059120540855664}} only a few academic studies have examined data from these tiplines: such studies include one on the 2019 Indian general election \citep{kazemi2022research} and one on the Brazilian general election \citep{hale2024analyzing}. Previous work has also focused on `public' WhatsApp groups \citep{garimella2018whatapp}. We add to this nascent literature by studying the 2021 Indian Assembly Election, focusing on the debunked claims and the overlap of users by using aggregated, anonymized data provided by fact-checkers.

Elections during COVID-19 happened with many uncertainties: about the unfolding medical situation, different rules \& regulations to control COVID-19, and statements made by political leaders during the campaign \citep{akbar2022political}. The same uncertainties are observed on social media platforms such as Twitter (now X) \citep{shahi2021exploratory} and Facebook \citep{arabaghatta2021covid}. Different fact-checking organizations came together to debunk misleading content on COVID-19 \citep{shahi2022amused}. 

The quinquennial Legislative Assembly Election held in 2021 marked an unprecedented year for India amid the worst public health emergency in the country, COVID-19.
The election period coincided with the delta wave (aka the second wave) of the COVID-19 pandemic in India. Beginning in March 2021, this devastating second wave led to shortages of vaccines, hospital beds, oxygen cylinders, and other medical supplies in large parts of the country \citep{SC2021}. Furthermore, by late April, India led the world in new and active cases. On 30 April 2021, India became the first country to report over 400,000 new cases in a 24-hour period \citep{BBC2021}. During this crucial period, many people in India turned to social media and messaging apps to share election and health-related information (and misinformation) \citep{shahi2022amused}.

In recent years, fact-checking has gained much momentum as a crucial tool in the fight against misinformation \citep{RAND2019}. Until recently, social media giants such as Meta and Twitter relied on independent third-party fact-checkers to fight misinformation on their platforms. Several fact-checking organizations also run tiplines on WhatsApp to discover possible misleading content.
Within India, the Ekta consortium\footnote{\url{https://ekta-facts.com/}} brought together six Indian fact-checking organizations to address misinformation in 2018 and started focusing on misinformation in early 2021  during assembly elections. Ekta covered four states and one union territory during the 2021 Indian assembly elections. Each of the six organizations operated a `tipline' on WhatsApp: they advertised their WhatsApp numbers and asked Indians to forward potentially misleading content for fact-checking.



In this study, we follow recent scholarship and define \emph{misinformation} as false or misleading information shared by users regardless of  whether there is an intent to deceive~\citep[e.g.,][]{quelle2025lost}. Within the Ekta consortium, misinformation tips are verified by fact-checkers manually based on factual information collected from credible sources, such as government portals, academics, or medical doctors. Fact-checkers assess the factual accuracy of these tips and provide ratings or verdicts based on the sources ~\citep{shahi2021exploratory}. If they are unable to verify `tips' from credible sources, they assign ratings/verdicts as `inconclusive' or `out of scope' (ref. Table~\ref{tbl:other}). We analyze claims or `tips' sent to Ekta tiplines during the 2021 Indian assembly elections. While tips were received in multiple data formats, including text, audio, images, and videos, only text-based tips were considered in this study. Further, fact-checked tips were collected in multiple languages, including Hindi, English, Bengali, Telugu, Marathi, Urdu, Tamil, and French. After data cleaning (see Sec.~\ref{sec:lang}), we analyze a dataset of 580 tips from 451 users in the three languages that each have more than 100 tips verified by human fact-checkers (English, Hindi, Telugu). We enrich this data by annotating the claims into three different categories (COVID-19, Election, and Other). We also normalize the ratings/verdicts obtained for these tips into four labels (\textit{false, partially false, misleading, and other}) as proposed in previous study~\citep{shahi2021exploratory} and explained in Section \ref{dat:cleaning}.  Table \ref{tbl:example-claims} shows some examples of claims received on the tiplines in different languages, and 
Table \ref{tbl:other} shows some examples of claims in the Other category.
The focus of this study is to answer the following research questions: 
\begin{description}
\item[RQ1] How do claims spread across different languages and fact-checking organizations during the 2021 assembly elections in India? 
\item[RQ2] How do fact-checked tips compare between different languages and across fact-checking organizations in India during the 2021 assembly elections? 
\item[RQ3] How quickly did fact-checking organizations in respond to user requests for checking potentially misleading content in India during the 2021 assembly elections? 
\end{description}

We examined the language of claims sent to WhatsApp tiplines and selected languages with at least 100 claims. The languages meeting this criteria were English, Hindi, and Telugu. We further translated non-English content into English using Google Translate and manually checked the translations. To answer RQ1, we plotted a timeline to analyze the flow of tips over time in different languages. We filtered the claims in three languages in terms of the ratings of the claims by fact-checkers. We also plotted word clouds to compare the top words appearing across languages. 

The elections happened during the peak of 2nd wave of COVID-19; so, we analyzed the claims qualitatively and annotated them in three different categories: claims related to COVID-19, Election, and Others. These categories provided a distribution of claim topics across different languages. To answer RQ2, we quantitatively analyzed the content overlap and presented the distribution of different categories of tips across three languages. In terms of tokens, we calculated the most frequent words for each language and compared across languages, creating a plot of frequent words across different languages. We also used hierarchal clustering to group claims from different languages having similar context. The content analysis of claims is discussed in Section~\ref{sec:content-overlap}.

To answer RQ3, we measured the time spent by fact-checkers to debunk claims to determine the speed of fact-checking organizations by computing the difference between the date of debunking and the date the claim was first received.  
We have normalized the ratings of fact-checked tips and found that tips are out of scope or lack sufficient information to be debunked in around 66\% of cases.
We also analyzed repeated users, who send claims across different tiplines, and grouped them by language and category.  Our user-level analysis is reported in Section~\ref{sec:user-analysis}. Our key contributions to this study are as follows:
\begin{itemize}
\item Our study is focused on an end-to-end encrypted platform, i.e., WhatsApp, during the 2021 Indian Assembly Elections.
\item We present manually annotated tipline data for claims during the election in three different languages, including the under-resourced language Telugu.
\item We provide different aspects of users and fact-checked responses in tiplines, which might provide useful knowledge for new fact-checkers setting up tiplines.
\end{itemize}

Our study illustrates claim detection on WhatsApp during a few weeks before and after the 2021 Indian assembly election, the speed of debunking claims, and how fact-checks reach users. Through an analysis of tipline metadata and content, we find a large overlap in the content across languages and organizations. We also find users who submit content in multiple languages. In contrast, no users submitted content that was ultimately fact-checked to multiple fact-checking organizations. This suggests each fact-checking organization cultivates a unique audience with their tipline. As such, it provides valuable insight into the role of WhatsApp tiplines in debunking claims during elections. We provide examples of an annotated list of tipline claims (tips) in Table \ref{tbl:example-claims}.




The remainder of this paper is structured as follows: Section \ref{sec:rw} reviews previous research and provides background information on the 2019 Indian Assembly Election, WhatsApp tiplines, and COVID-19 in India. Section \ref{sec:research} details the dataset and methodology, outlining the various steps involved in the study. Section \ref{sec:r} presents the results, while Section \ref{sec:d} discusses key observations and provides recommendations for practitioners. Finally, Section \ref{sec:c} concludes the paper and suggests future research directions.




\section{Related Work \& Background}
\label{sec:rw}
In this section, we discuss the previous work analyzing the misinformation in encrypted platforms and Indian elections and provide background on the 2021 Indian Assembly Elections and WhatsApp tiplines.
\subsection{Misinformation and Elections}

Misinformation is wider than social media, and misleading information has likely existed throughout most of human history \citep{burkhardt2017history}. While other terms such as fake news, disinformation, and malinformation have been used to describe various aspects of false or misleading content, we use the term misinformation as an umbrella term to refer to any content that is false or misleading regardless of the intent behind it in line with other recent scholarship on the topic \citep[e.g.,][]{altay2023misinformation}. In our study, we examine content fact-checked by organizations that follow the International Fact-checking Network Code of Principles.\footnote{\url{https://www.ifcncodeofprinciples.poynter.org/}}

The spread of misinformation gained prominence with the rise of social media, drawing significant scholarly attention, especially following the 2016 U.S.\ Presidential election \citep{bovet2019influence}, the 2017 French presidential election \citep{howard2017junk}, and other nation elections. 
Research has examined aspects such as bots \citep{graham2023bots} and echo-chambers \citep{shane2022rise} on different platforms (e.g., Twitter, YouTube). \citet{neyazi20212019} assert that misinformation played a role in shaping the the 2019 Indian general election. Misinformation during the election is not limited to social media such as Twitter and Facebook. Rather, it extends to end-to-end encrypted platforms such as WhatsApp, Signal, and Telegram, which also play an important role in disseminating misinformation during elections \citep{shahi2022amused}. This is especially true of WhatsApp in its largest markets: India \citep{kazemi2022research} and Brazil \citep{machado2019study}. 

\citet{jakesch2021trend} present the use of manipulated campaigns in the form of mobilization messages on WhatsApp with lists of pre-written tweets. They find hybrid cross-platform network generated hundreds of nationwide trends on Twitter and volunteer participation in political groups lead to a high reach for election campaigns.

\subsection{Misinformation on Encrypted Platforms}

Previously, \citet{shahi2023exploratory} showed that WhatsApp is known for the spread of misinformation in India. Notably, much content is shared to WhatsApp tiplines before it appears in datasets of large, public groups on the platform \citep{kazemi2024human}. 
\citet{seelam2024fact} conducted a user study with multiple fact-checking organizations and found that they used everyday technologies such as WhatsApp tiplines and bots to receive claims and
share fact-checks, with a focus on rural users because of the ubiquity and simplicity of WhatsApp. 
\citet{kazemi2022research} found unique misinformation on WhatsApp that was not present on Twitter during the 2019 Indian general elections, underscoring the significance of encrypted platforms in this context. 
Similarly, using large public groups, \citet{machado2019study} found misinformation spread on WhatsApp during the 2019 presidential election in Brazil and made heavy use of visual metaphors.
During the Nigerian election, \citet{machado2019study} found that the political parties sought to use encrypted apps for election manipulation. It is difficult to estimate what percentage of content on end-to-end encrypted platforms is false or misleading; however, \citet{garimella2020images} suggest up to 13\% of images shared on large, public WhatsApp groups during the 2019 Indian general election were false or misleading. \citet{reis2020can} collected an image dataset from public WhatsApp groups during elections in India and Brazil to analyze the spread of misinformation, even after debunking, misinformation images were circulated within public groups. The authors proposed an architecture for device-level fact-checking of debunked images using hash-based verification. Expanding on this work, the present study investigates on-demand fact-checking of suspicious claims by facilitating direct interaction between users and fact-checkers via tiplines.


To counter misinformation, WhatsApp introduced various measures including labeling highly-forwarded messages, limiting the size of groups, and limiting forwarding behavior \citep{shahi2023exploratory}. WhatsApp has also encouraged third-party fact-checkers to establish tiplines. 
Following that, \citep{pasquetto2022social} studied the role of WhatsApp in correcting misinformation through the resharing of debunks and found users reshared debunks at a large rate when they received them from users close to them (strong ties).
Thus adoption of content verification techniques even by a small group of users could reduce misinformation in India's WhatsApp networks \citep{jain2021misinformation}.



\subsection{2021 Indian Assembly Election \& COVID-19}\label{sec:background-election-covid}
India is composed of 28 states and 8 ``union territories''. In India, the Election Commission of India (ECI)\footnote{\url{https://eci.gov.in/}} conducts the assembly election every five years to elect members of the legislative assembly within a state or union territory. Later, elected members choose the chief minister of the region. Almost every year, some Indian states have their assembly elections. In 2021, assembly elections were conducted in four states (Assam, Kerela, Tamil Nadu, and West Bengal) and one union territory (Puducherry). The assembly elections are conducted across multiple phases to conduct a fair election. Each phase includes a list of assembly constituency and a separate polling date. Still, the results are declared on the same day. Each regions have own nomination dates, however, the first nominations for political candidates started on 9 March 2021 in West Bengal \& Assam, and voting occurred from the end of March to the end of April with the results declared on 2 May 2021 for all regions. Different issues (e.g., COVID-19) were discussed during the elections. In this study, we focus on three months---March, April, and May 2021---to study misinformation claims debunked by Ekta consortium members. 
In India, the second wave of COVID-19 started in March 2021 and resulted in a shortage of medical supplies nationwide. The second wave evolved at a phenomenal speed compared to the first wave driven by the Delta variant. In India, approximately 331,895 people died in the second wave of COVID-19 (through 31 May 2021).\footnote{\url{https://coronavirus.jhu.edu/region/india}}  At the same time, the ECI announced the assembly elections would have physical, in-person voting. During the campaigns, COVID-19 cases rose rapidly in the regions with elections compared to other states without elections \citep{manik2022effect}, and the ECI imposed several restrictions overall during the election periods to ensure the implementation of preventative measures by the candidates while they were campaigning. Due to these, multiple claim requests were submitted to fact-checking organizations for COVID-19 and election-related issues. 


\subsection{WhatsApp Tiplines}
In early 2019, PROTO
(an India-based media skilling startup) hosted the first official WhatsApp fact-checking tipline in the Checkpoint project using Meedan's open-source software Check\footnote{\url{https://tinyurl.com/meedan1}} during the 2019 Indian general elections \citep{kazemi2022research}.
We call the messages from WhatsApp users tips, and they have content in multiple data formats including text, image, video, audio, and links. A tip can be further classified as a claim if it contains a fact-check-worthy claim \citep{hassan2017toward}.

At the end of 2019, several Indian fact-checking organizations began operating always-on tiplines that were not specific to any election or other event. Later a subset of organizations operating tiplines in India formed the Ekta consortium.

\begin{figure}[]
    \centering
    \includegraphics[height=10cm]{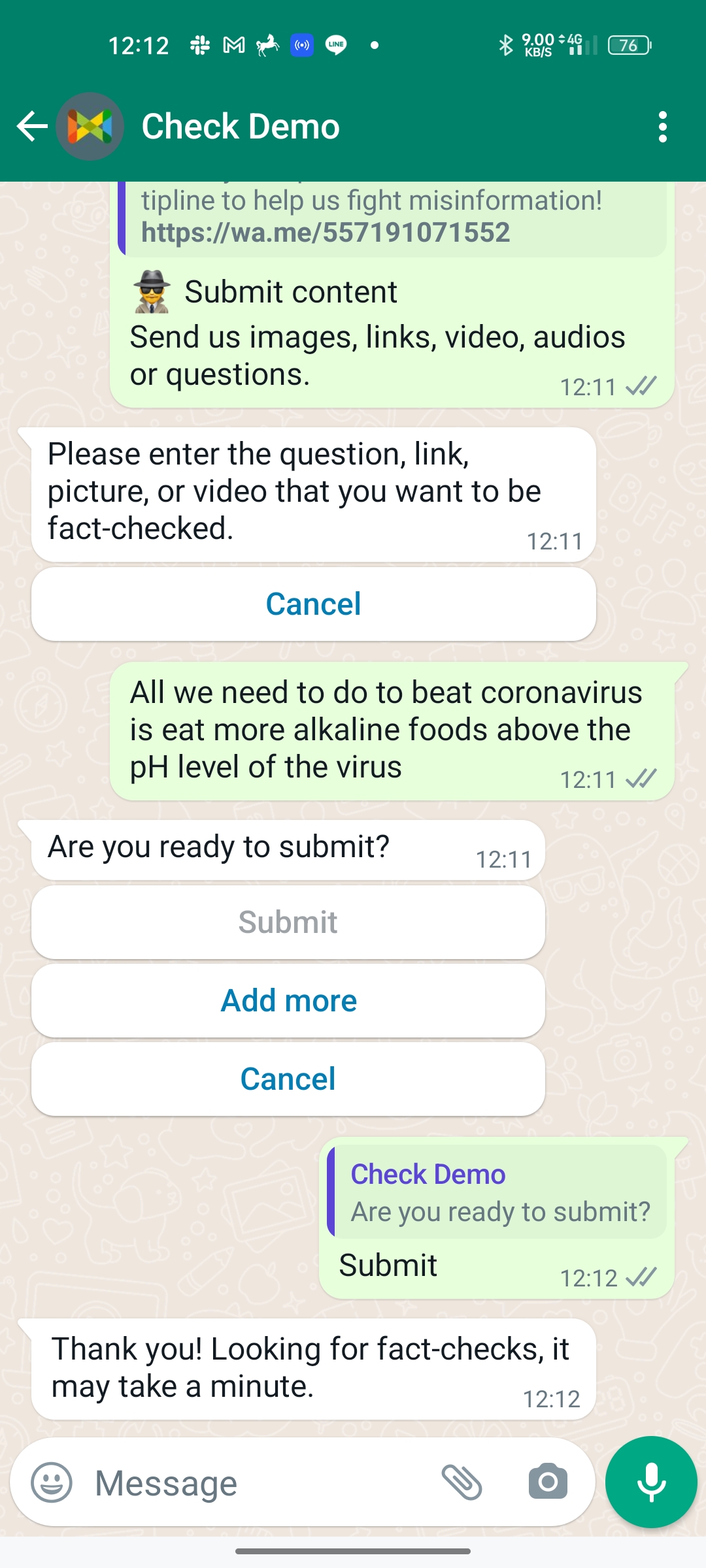}\hspace{.05\textwidth}%
    \includegraphics[height=10cm]{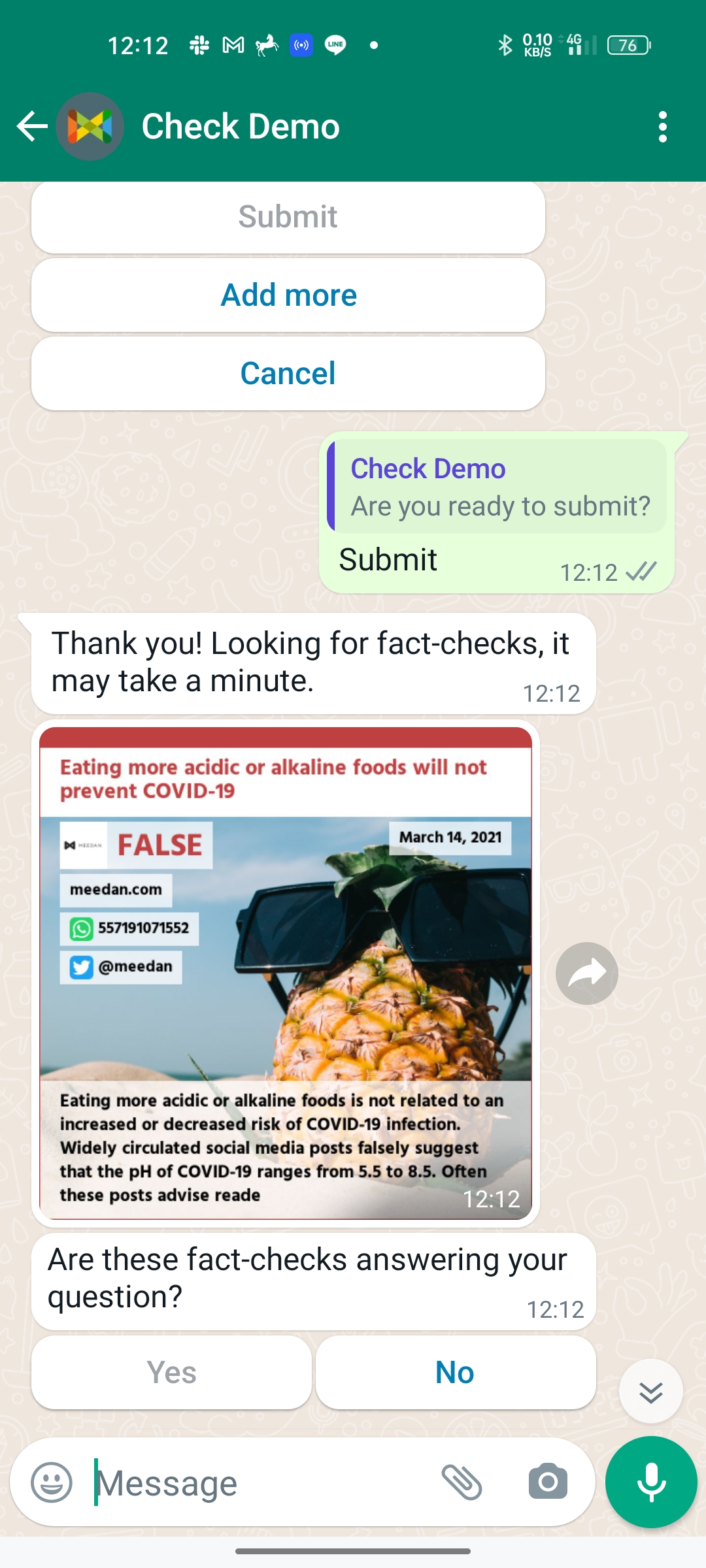}%
    \caption{An example of using WhatsApp to get a verdict for a claim}
    \label{fig:tipline}
\end{figure}

Screenshots of a demo WhatsApp tipline are shown in Figure \ref{fig:tipline}. When users see suspicious stories or images, they can send them to a WhatsApp tipline (a mobile number with a WhatsApp account). In this example, the demo fact-checking organization is asked about alkaline foods and COVID-19, and a fact-check is returned immediately as the question has been fact-checked previously.  

\begin{figure}[!htbp]
    \centering
    \includegraphics[width=.56\textwidth]{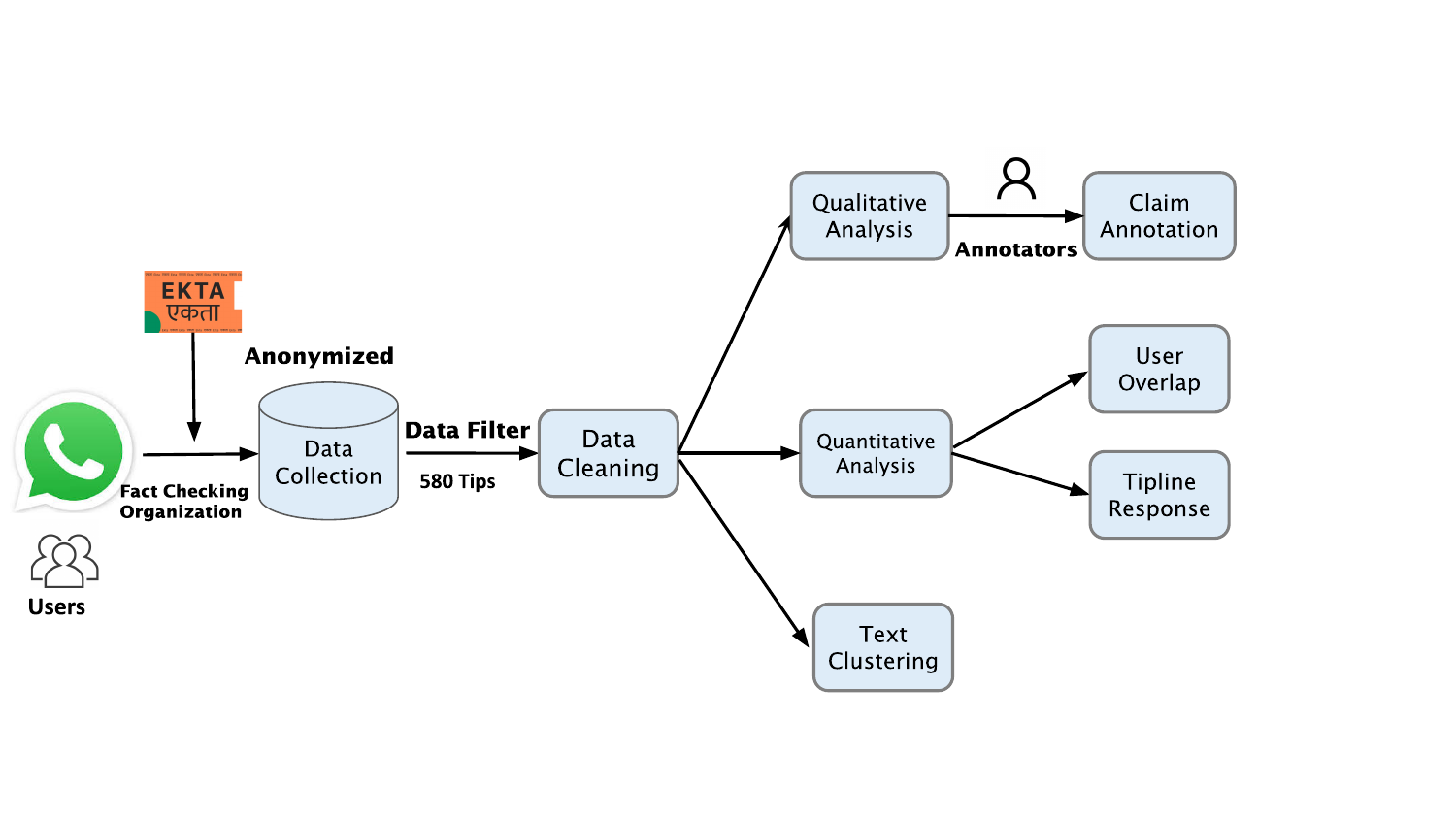}
        \caption{Overview of research methodology}
    \label{method}
\end{figure}


\section{Data and Methods}
\label{sec:research}

This section discusses the steps in analyzing the tipline data for the 2021 Indian Assembly Elections. The process of analysis is shown in the research diagram in Figure \ref{method}.

\subsection{Data Collection}
\label{sec:data}

We collected anonymous and aggregated data from Ekta members on an opt-in basis for this study. The data contains the text of the tip requested for fact-checking, the date/time the tip was received, and the linked fact-check. We received no information about the users submitting the tips beyond an anonymous identifier that allowed us to check if the same user submitted content to multiple tiplines. As part of our agreement, we agreed to only examine the data in aggregate and not identify any specific fact-checking organizations.
The data covered two weeks before the first nomination (9 March 2021) and four weeks after the result declaration (2 May 2021), i.e., 1st March 2021 to 31st May 2021. 
Of the text tips received during this time frame, we selected 950 textual claims that were fact-checked in order to focus specifically on claims known to be misinformation.  A detailed description of all tips after cleaning, translation, and qualitative analysis is given in Table \ref{data_description}.
 

\subsection{Language Identification}
\label{sec:lang}

As the elections happened in five states with different regional languages, we first identified the language of the claims submitted to the tiplines using Python library (pycld3). pycld3\footnote{\url{https://pypi.org/project/pycld3/}} is a Python implementation of CLD3,\footnote{\url{https://github.com/google/cld3}} which is a neural network model for language identification using ngrams and capable of identifying multiple languages. 
Through random checks, we found that some errors occurred when the confidence score was below 0.90. Therefore, all texts with a confidence score below 0.90 were reexamined by a language expert, who confirmed the detected languages. Ultimately, the researcher manually corrected the language of 36 tips.

\subsection{Data Cleaning \& Preprocessing}
\label{dat:cleaning}

\begin{figure}[!htbp]
    \centering
    \includegraphics[width=.49\textwidth]{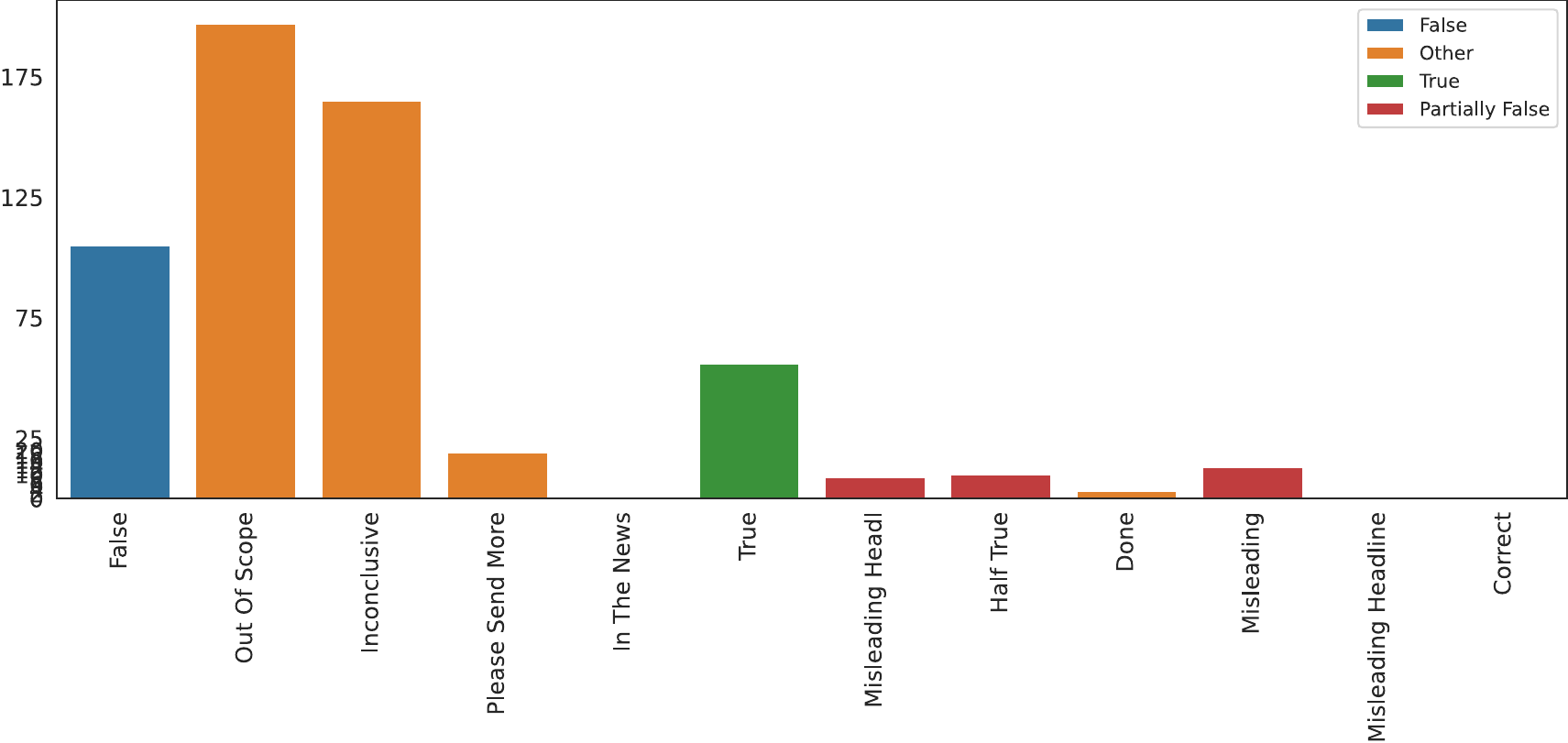}
        \caption{Normalization of ratings/verdicts of tipline claims (X-axis shows name of ratings/verdicts and Y-axis shows thier counts)}
    \label{fig:ratings}
\end{figure}

The data cleaning and preprocessing include several steps. After filtering claims, we removed numbers and URLs from the claims. 
We removed stopwords, emojis, special characters, and short words with less than three characters using NLTK \citep{bird2006nltk}. We also removed messages with less than five words 
before preprocessing. After preprocessing, some claims had less than five words. For example, \textit{``Sir, this is fake, are not"} was cleaned to ``fake'' (one word only). Most such cases have a rating/verdict of Other. 
We merged the Hindi (hi) and Hindi English code switch claims (hi-latin) into Hindi claims, and then we translated all claims into English using Google Translate. The mean length of claims is 150 words. We found that the 580 claims have 12 different ratings/verdicts given by fact-checking organizations, including `Misleading' and `Correct'. As this creates sparse groupings, we mapped the ratings/verdicts into four categories: false, partially false, true, and others, using the mapping rules discussed by \citet{shahi2021exploratory}. 
This only involved mapping existing ratings/verdicts and not re-evaluating the original content. The Other rating includes \textit{out of scope, inconclusive, please send more}, and the distribution of ratings/verdicts varies across the three languages we study. The normalization of 12 different rating into 4 four categories is shown in Figure \ref{fig:ratings}. Original ratings such as ``in the news'', ``misleading headline'' and ``correct'' appears only once and ``done" appears three times.
The obtained class distribution is shown in Figure~\ref{fig:category}a. 

\begin{table}[!htbp]
\caption{Description of Tipline Dataset}
\label{tab1}
\renewcommand{\arraystretch}{1.2} 
\setlength{\tabcolsep}{6pt} 
\begin{tabular}{@{}p{3.5cm} p{0.7cm} p{0.5cm} p{0.6cm} p{0.7cm}@{}}
\toprule
\textbf{Type} & \textbf{English} & \textbf{Hindi} & \textbf{Telugu} & \textbf{Overall} \\
\midrule
Claims & 271 & 199 & 110 & 580 \\
Claim length (words) & 131 & 160 & 179 & 150 \\
Users & 237 & 173 & 61 & 451 \\
Time to Debunk (Days) & 2.15 & 2.98 & 3.2 & 2.92 \\
\midrule
\textbf{Ratings} &  &  &  &  \\
\midrule
False & 58 & 26 & 21 & 105 \\
Partially False & 16 & 10 & 7 & 33 \\
True & 23 & 16 & 19 & 58 \\
Others & 174 & 147 & 63 & 384 \\
\midrule
\textbf{Category} &  &  &  &  \\
\midrule
COVID-19 & 118 & 56 & 41 & 215 \\
Election & 35 & 67 & 41 & 143 \\
Others & 118 & 76 & 28 & 222 \\
\bottomrule
\end{tabular}
\label{data_description}
\end{table}

\subsection{Qualitative Analysis}

The 2021 Indian Assembly Elections were conducted during the 2nd wave of COVID-19. There were several uncertainties and chaotic situations in different states of India. Also due to a lack of proper clarification, misinformation was circulated about COVID-19 and the elections.  We qualitatively identify the topics of the tipline claims to know more about the context of the claims.


We annotated the claims into one of three categories---election, COVID-19, and others---based on an inductive, grounded theory approach. On our first pass through the data, we identified that most claims were related to either COVID-19 or the elections. We placed any claims unrelated to either of these into the other category. The description of the three categories are given below.

\begin{itemize}
    \item COVID-19: This category includes all claims that ask about COVID-19, including vaccination, cures, and masking. We also include the claims about medicines related to COVID-19. For example, ``can inhaling steam help in fighting COVID-19” 
    \item Election: This category includes the claims that ask about election-related issues. Claims include the issues from the 2021 Indian general election and negligence of COVID-19 during the election. When a claim is about COVID-19 in relation to the election or politicians, it is categorized as an election-related claim. The claims in this category also include national issues affecting other states. For example, \textit{``whether Indian politician Mamata Banerjee faked a leg injury''.\footnote{\url{https://tinyurl.com/yrz5suss}}} 
    
    \item Other: This category accumulates all other claims which do not fit in any of the above two types. It includes claims on international issues, asking for email IDs, and fact-checking reminders. An example of a claim is, ``sir police got rid of money and made false cases"
\end{itemize}

Annotation was performed by two researchers working in the area of fact-checking. Based on the descriptions, we have annotated the translated text of 580 claims by one annotator. Another annotator randomly sampled 100 claims and annotated them. We computed the intercoder reliability using Cohen's kappa and found it as 0.86, showing that both annotators agree with each other in most cases. From the qualitative coding of claims, we found 215 COVID-19, 143 Election, and 222 Other claims in three different languages. Some examples of claims in three categories are shown in Table \ref{tbl:example-claims} and the category distribution across languages is shown in Figure \ref{fig:category}b. We have provided examples in different languages with their translations, ratings/verdicts, and categories of tips. The tips of COVID-19 show the uncertainty of the lockdown situation and the impact of different variants of coronavirus. The political category includes a tip about Modi's rallies, and Other category tips are about the issue such as alleged fraud.

\begin{table*}[!htbp]
    \centering
    \caption{Examples of tipline claims in different languages}
    \includegraphics[width=.99\textwidth]{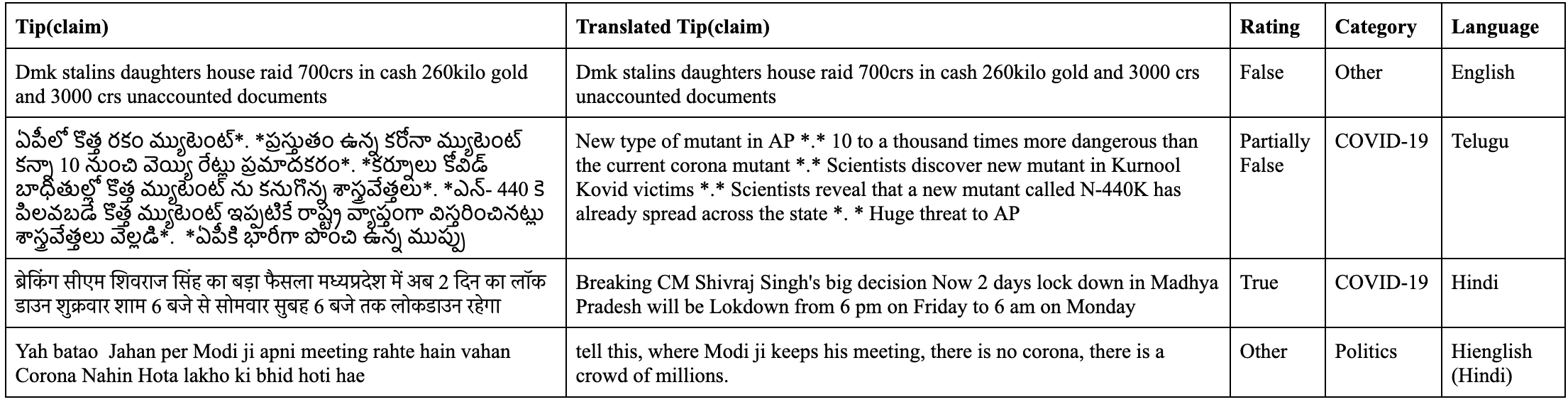}
    \label{tbl:example-claims}
\end{table*}

\subsection{Timeline of Claims}
The Indian assembly elections were conducted for 55 days (between the first date of nomination and result declaration), and we analyzed the data for 92 days. In addition, the second wave of COVID-19 peaked during this time. So, we plotted the timeline analysis to visualize the number of fact-checked text claim requests coming each day and peak claims during a particular time. The timeline analysis shows the flow of incoming tips during the election phase.

\subsection{Content overlap}\label{sec:content-overlap}
In order to understand the content overlap across languages and organizations (RQ2), we performed a qualitative analysis of the claims that were fact-checked by different fact-checking organizations in different languages. We also consider the volume of claims per language over time and the overlap in the most frequent words by finding common words across different languages.




\begin{table*}[!htbp]
    \centering
    \caption{Example tipline claims in different languages in Other Category}
    \includegraphics[width=.99\textwidth]{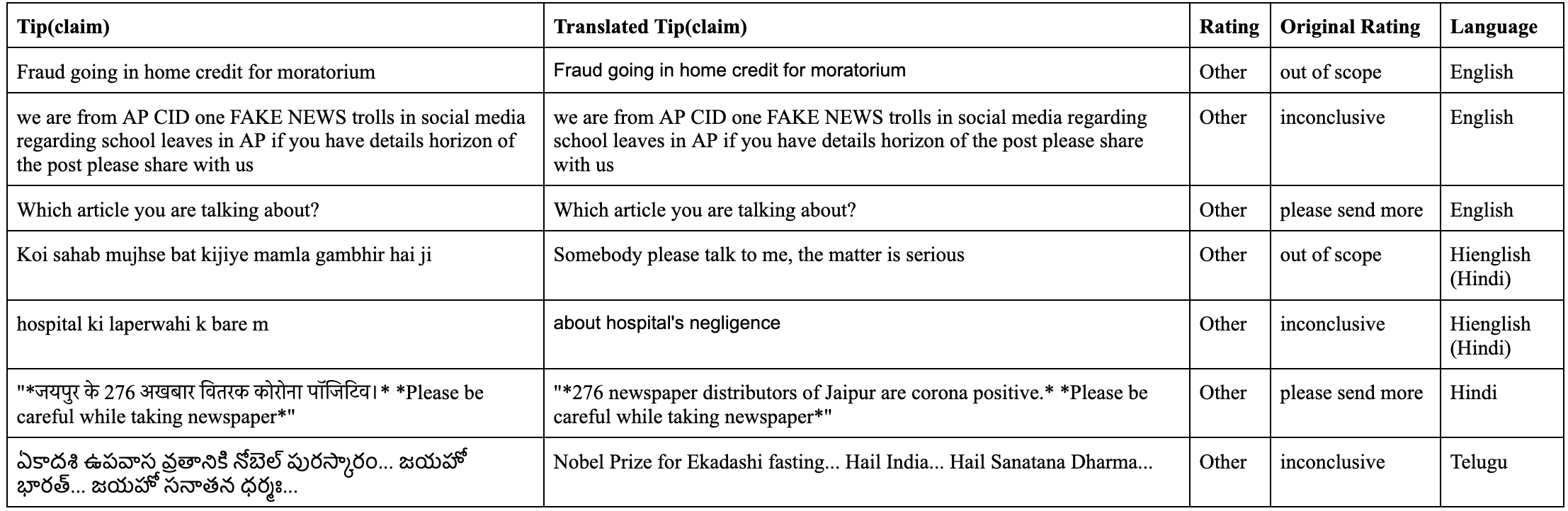}
    \label{tbl:other}
\end{table*}

\subsubsection{Frequent Words}
We computed the frequent terms used in each language. To visualize the frequent terms in each language, we used word clouds which give a unique insight into claims in different languages and help to visualize the most frequent terms. We used the translated claims in English to generate the word cloud to present keywords and their frequencies. Apart from visualizing the frequent words in each language, we also identified the common words in each pair of languages and among all three languages, i.e., Hindi--English, English--Telugu and Telugu--Hindi and Hindi--English--Telugu.



\subsection{Semantic Text Clustering}
Text clustering is an approach that helps organize a large corpus into smaller groups called clusters. Several practices have emerged for vector representation of textual data, including TF-IDF and word embeddings (e.g., Word2Vec, Glove). Embeddings help us to compute the semantic similarity between two words efficiently. With the growth of the language model, several models are used for text clustering. \citet{abbasi2021tourism} used semantic clustering to find tourism recommendations based on user reviews. Semantic text similarity has also been used for matching the similar claims \citep{kazemi2022research}. Fact-Cred used the RoBERTA-based semantic similarity for the semi-automatic framework for misinformation detection \citep{martin2022facter}. 

We embed tips using the BERT-based sentence transformer Indian XLM-R model for text clustering \citep{kazemi2022research}. We used the \textit{SciPy\footnote{\url{www.scipy.org/}}} library to implement hierarchical agglomerative text clustering \citep{Shalizi2009} to find the different groups. 
We used the preprocessed data as mentioned in Section \ref{dat:cleaning}. 
The clustering is initialized by assigning each claim to a unique cluster and then merging clusters until one root is left. We defined a function which takes the sentence embedding, computes the cosine similarity, and forms a condensed distance matrix using SciPy's ward function. 
The claims in each cluster represent context from original claims with similar tokens. 



\subsection{User Overlap \& Speed of Fact-checking}\label{sec:user-analysis}

We analyze the overlap of users across languages and organizations to answer RQ3. For each claim request sent to a tipline, we have a timestamp and pseudonymous users' ID. We analyze these data to understand 
the number of claims from the same user and in multiple languages. We also check whether the same user sent claims to multiple fact-checking organizations.  

We analyze the time required by fact-checkers to debunk a claim. The speed of fact-checking is important because the delay in providing the verdict helps the propagation of the misinformation~\citep{shahi2021exploratory}. Also, fact-checking organizations get multiple requests for the same message to debunk the claim, increasing their workload. To compute the speed, we calculated the difference between the date the fact-check was published and the date the claim was first submitted to the tipline.

\section{Results}
\label{sec:r}

In this section, we discuss the results obtained from content analysis, focusing on the tips and the overlap among users who sent tips to be debunked.

\begin{figure*}[!htbp]
    \centering
    \includegraphics[width=0.99
    \textwidth]{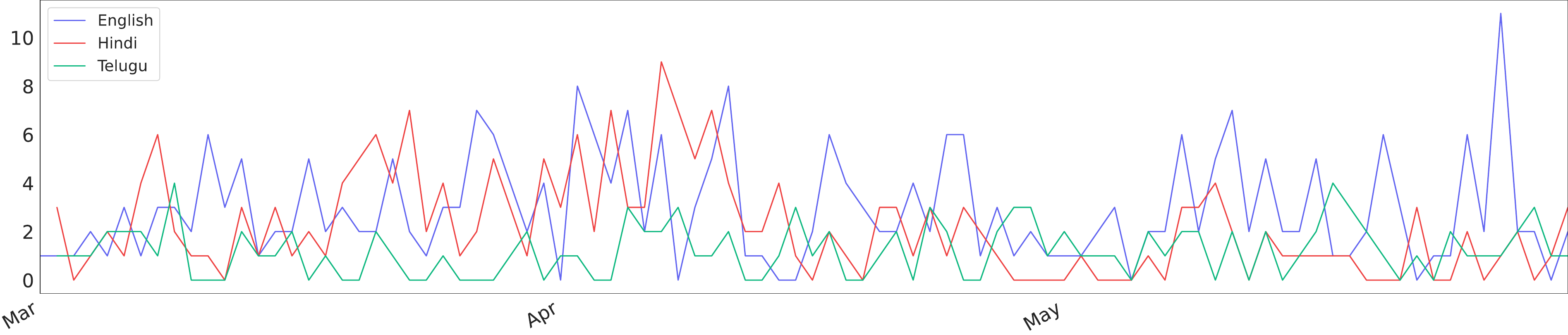}
    \caption{Timeline of number of textual fact-checked claims across three languages from March-May 2021}
    \label{fig:timeline}
\end{figure*}

\begin{figure}[h!]
    \subfloat[Verdict(Fact-checkers)\label{Verdict(Fact-checkers)}]{%
      \includegraphics[width=0.49\textwidth]{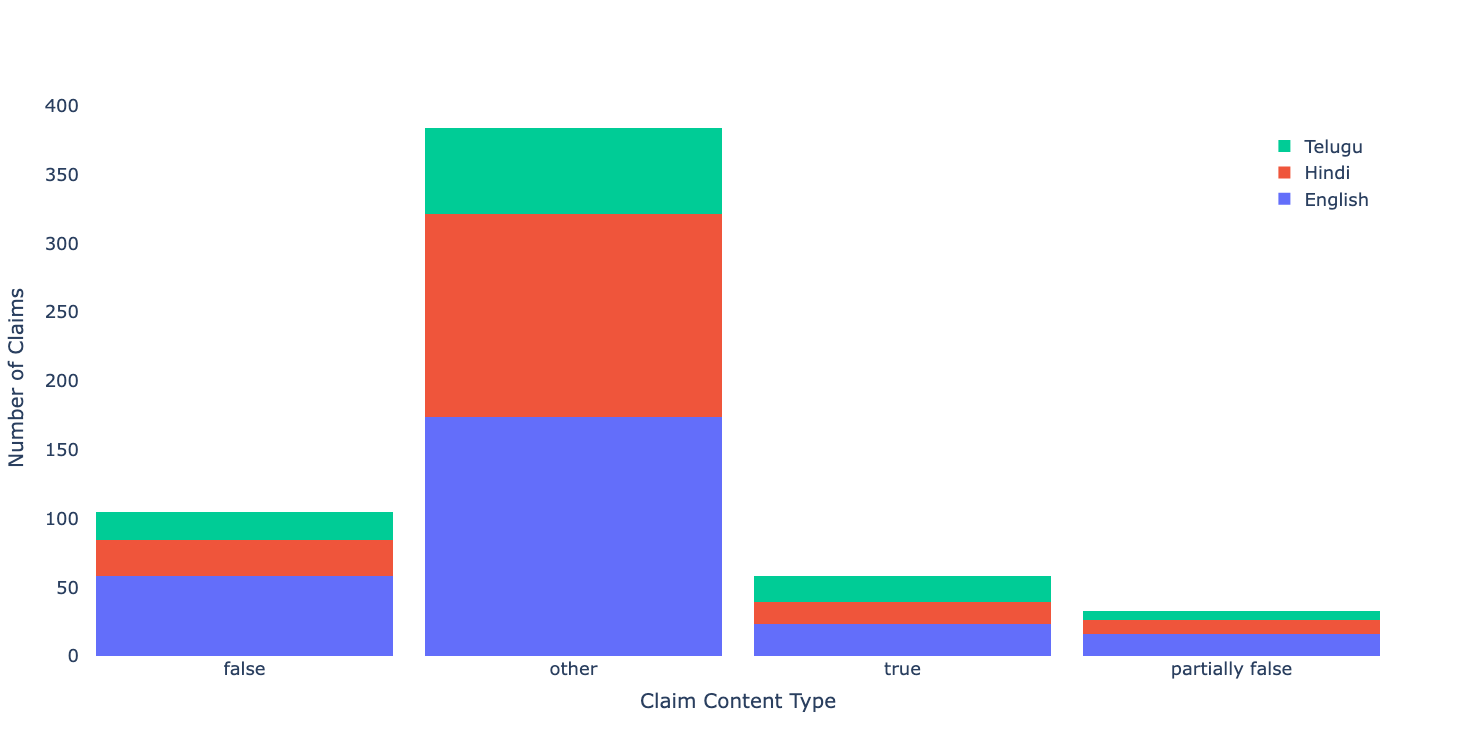}
    }
    \hfill
    \subfloat[Category(Annotated)\label{Category(Annotated)}]{%
      \includegraphics[width=0.49\textwidth]{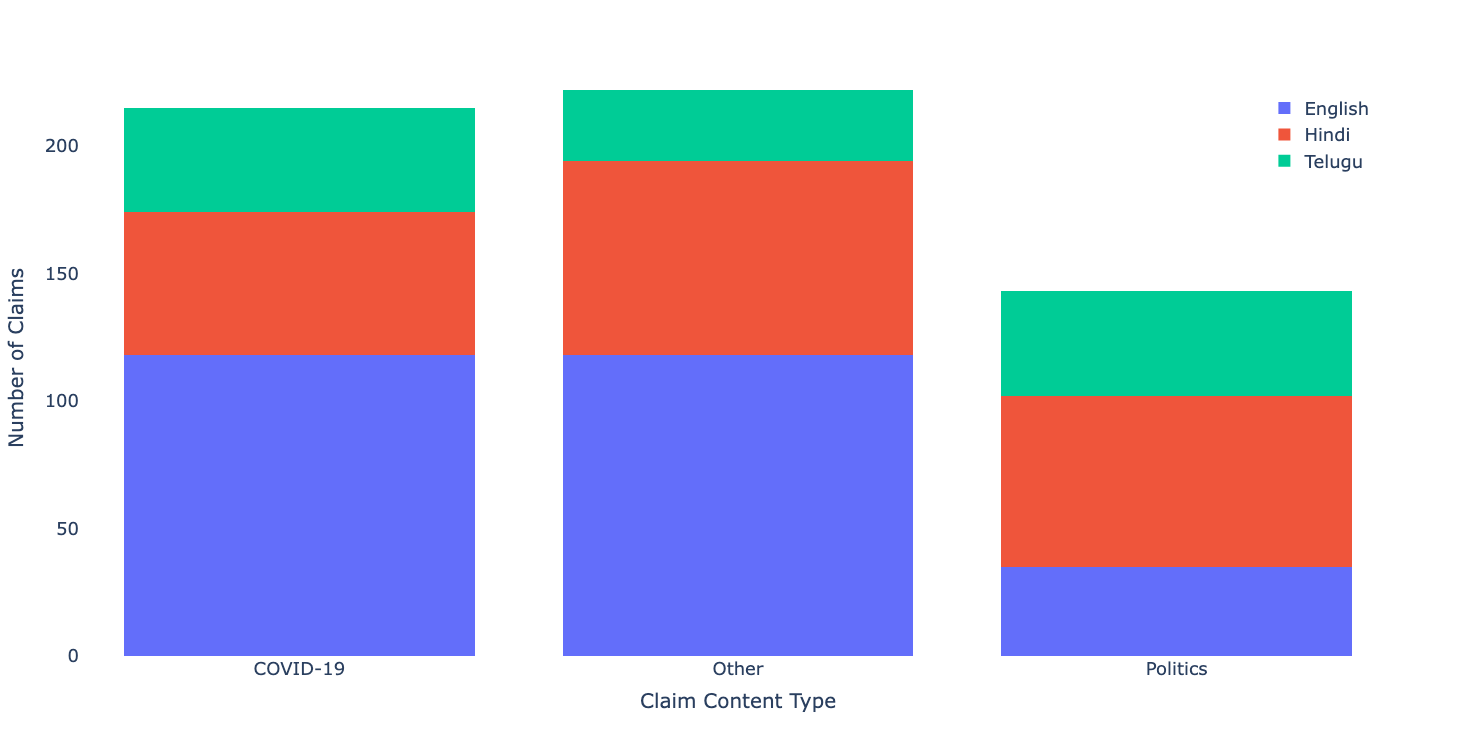}
    }
 
    \caption{Distribution of claims in different languages}
    \label{fig:category}
  \end{figure}

We start by answering RQ1. First, we plotted timelines of the tips that were fact-checked by human fact-checkers in the three languages. The timelines of tips in different languages are shown in Figure~\ref{fig:timeline}. The number of requests increased in March and then peaked in mid-April, which was the final round of voting in all regions. Then, it starts decreasing towards the end of April. There were not so many requests during and immediately after the declaration of elections' results. From the second week of May, there was an increase in requests, particularly in English. Fact-checkers provide different labels to debunked claims: we provided normalized labels into four categories, and their distribution is shown in Figure \ref{fig:category}a.


\begin{figure*}[h]
    \subfloat[English\label{English}]{%
      \includegraphics[width=0.33\textwidth]{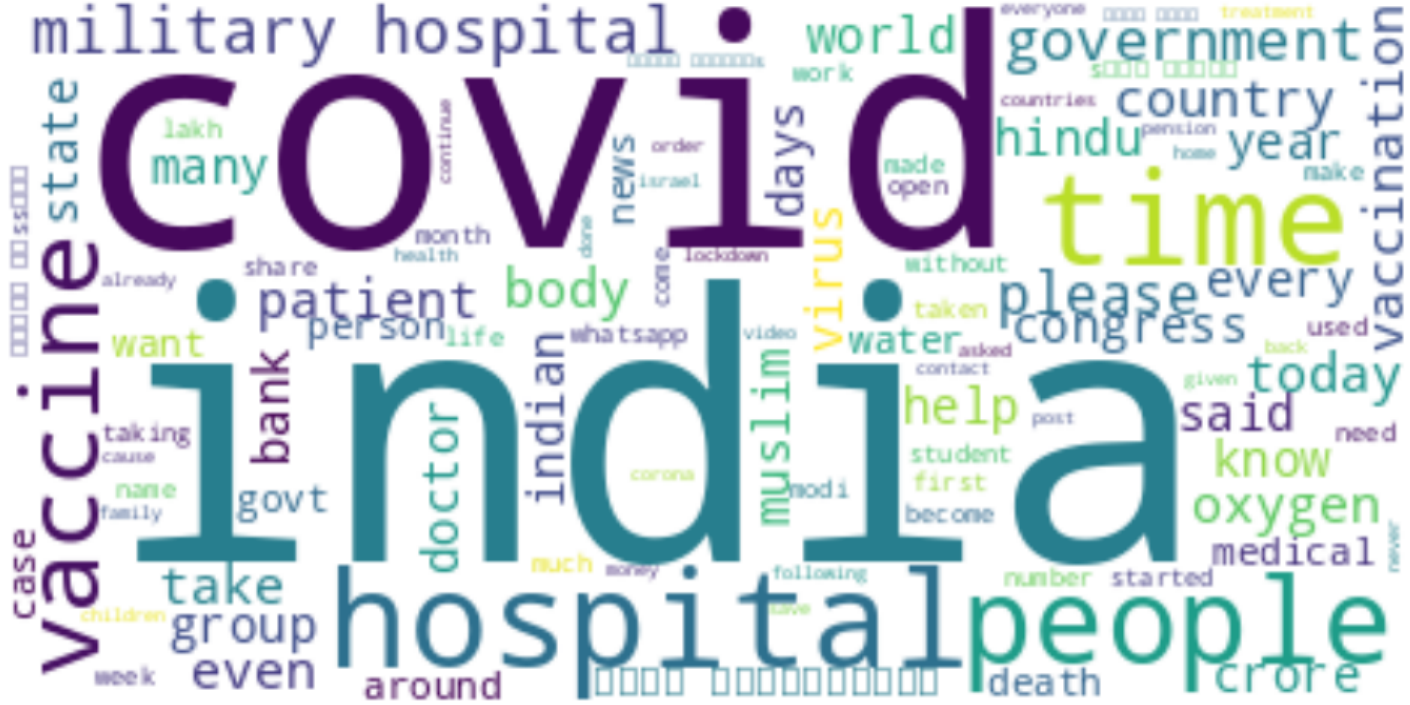}
     }
    \subfloat[Hindi\label{Hindi}]{%
      \includegraphics[width=0.33\textwidth]{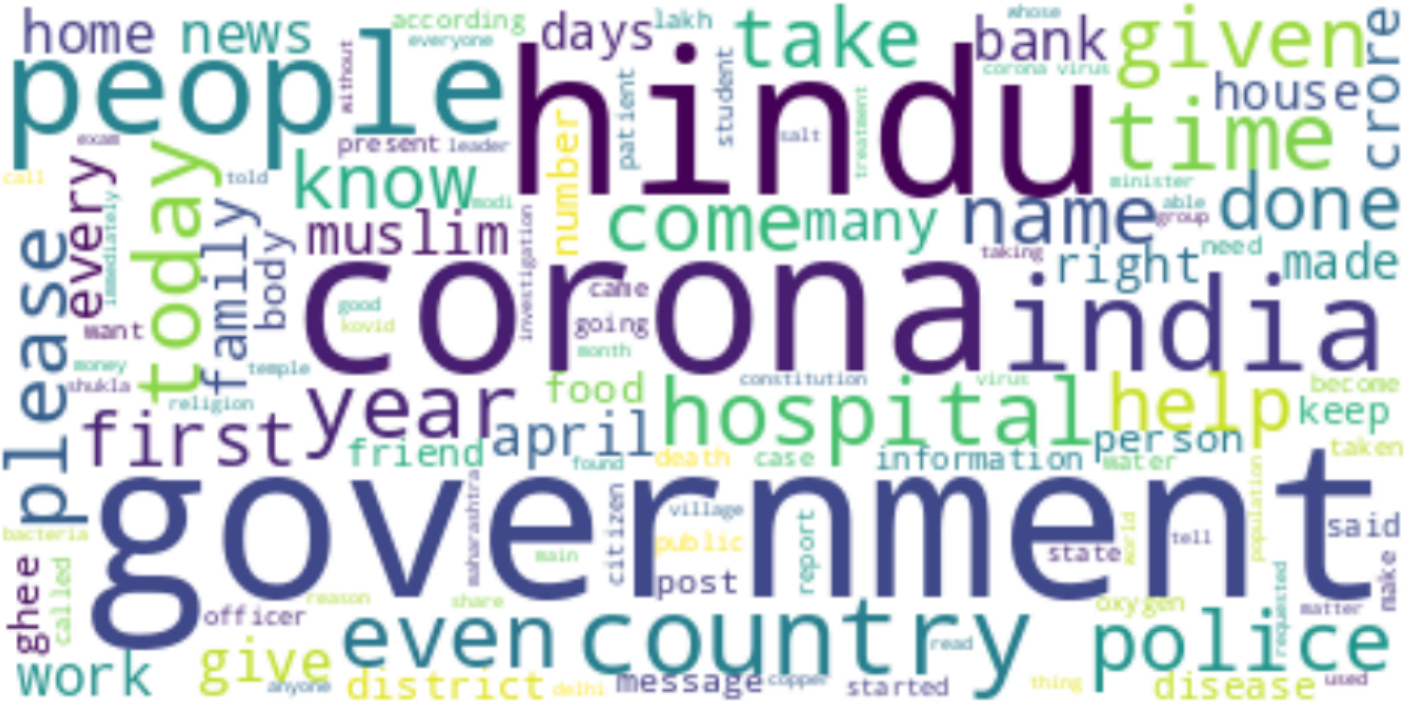}
    }
    \subfloat[Telugu\label{Telugu}]{%
      \includegraphics[width=0.33\textwidth]{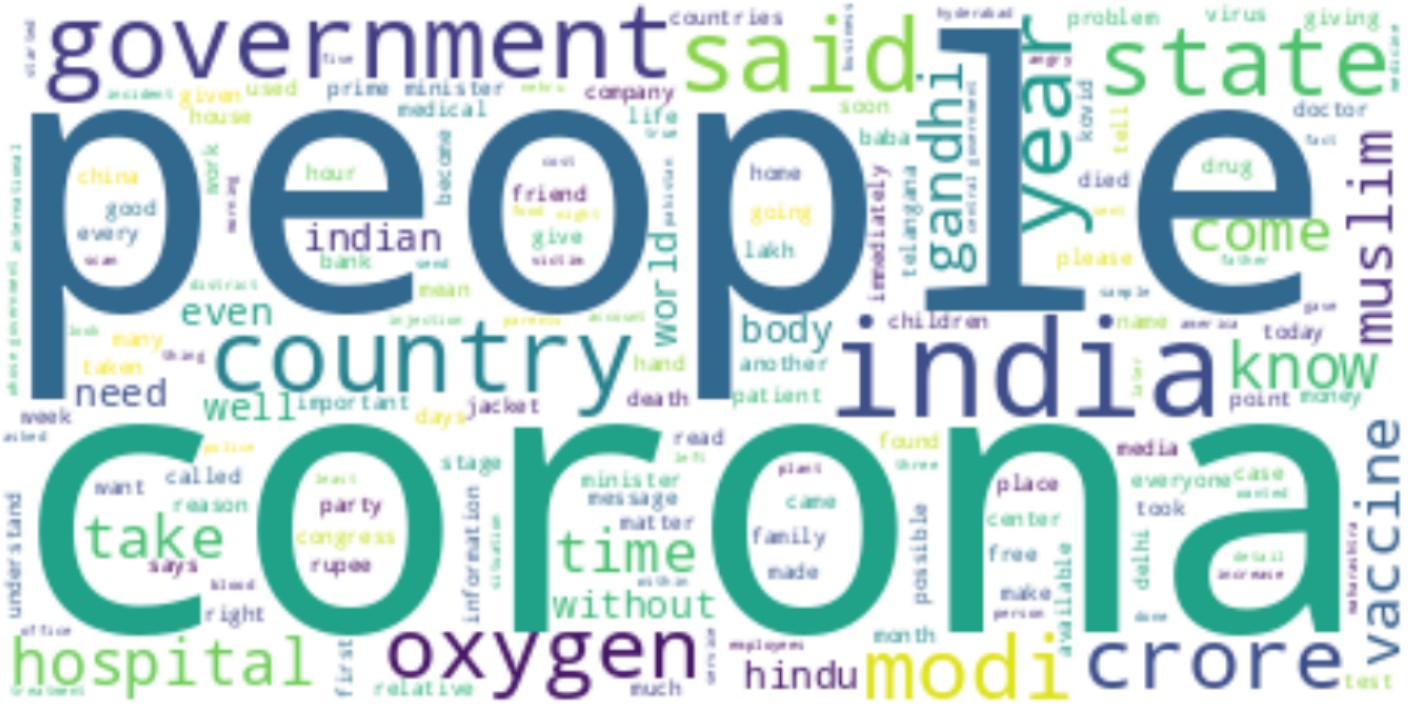}
    }
      \hfill
    \subfloat[Hindi-English\label{Hindi-English-Word-Cloud}]{%
      \includegraphics[width=0.33\textwidth]{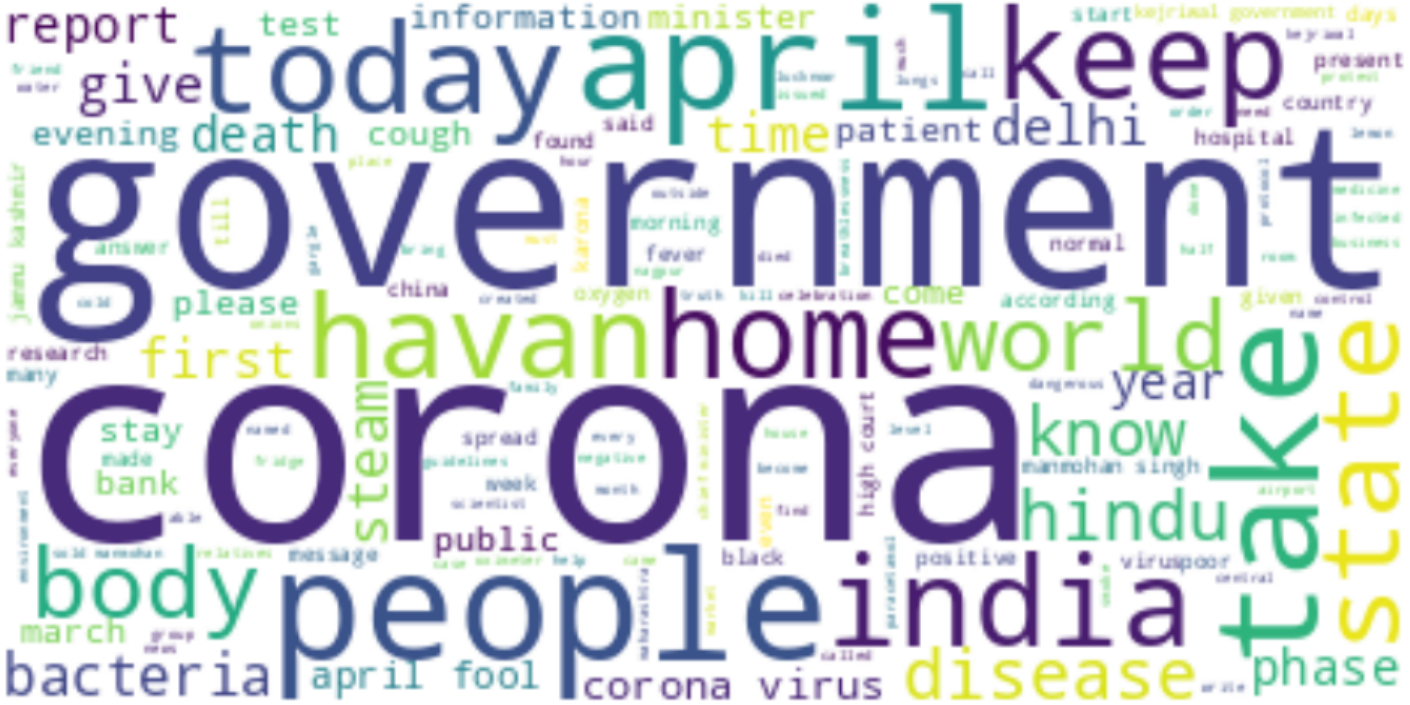}
    }
    \subfloat[Telugu-Hindi\label{Telugu-Hindi-Word-Cloud}]{%
      \includegraphics[width=0.33\textwidth]{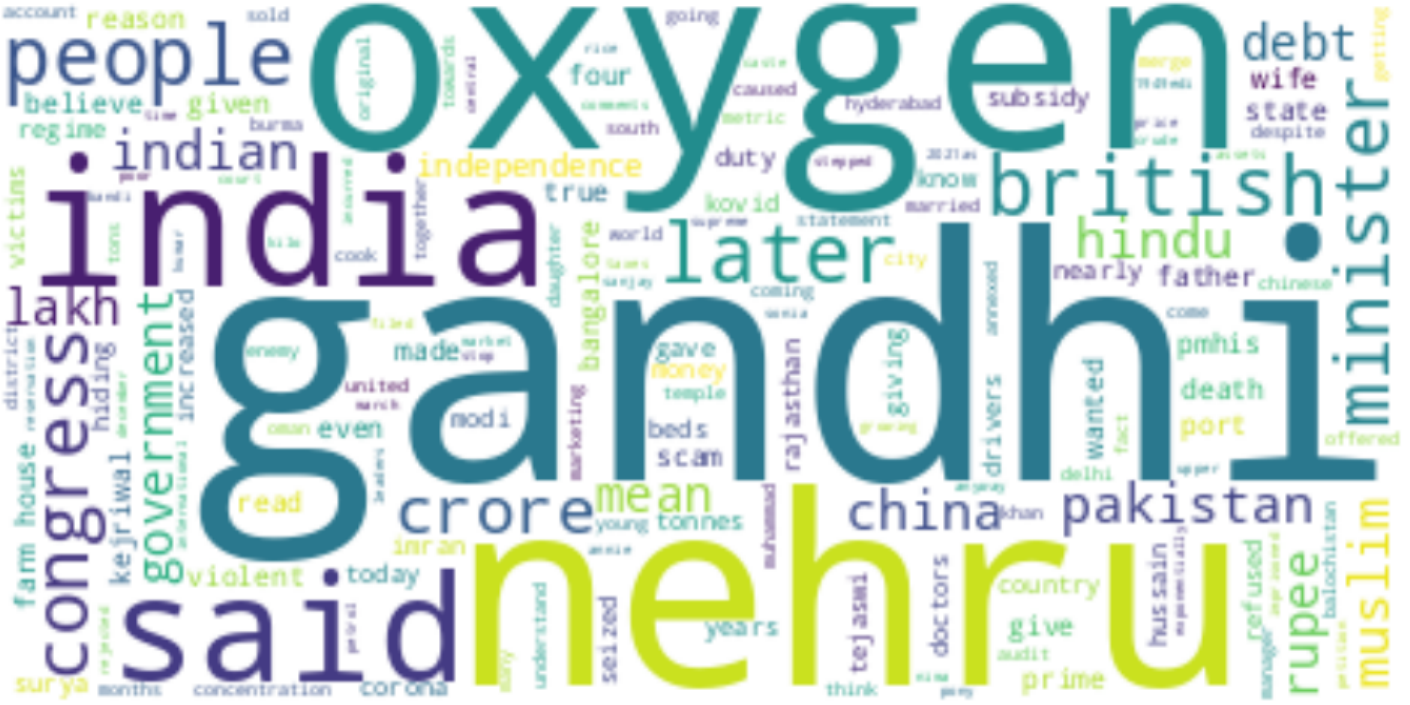}
    }
    \subfloat[Telugu-English\label{English-Telugu-Word-Cloud}]{%
      \includegraphics[width=0.33\textwidth]{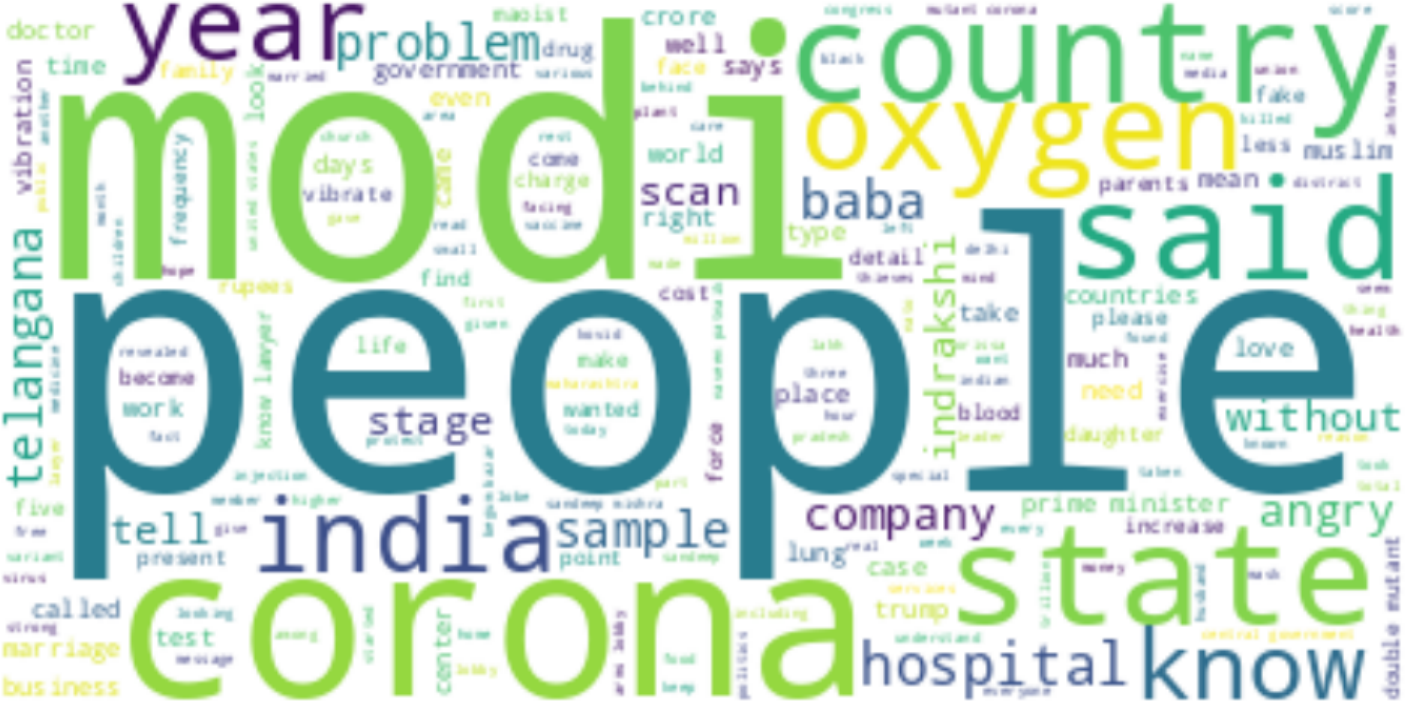}
    }
    \caption{Word clouds present the top words of mono and bilingual claims}
    \label{fig:wordclouds:perlanguage}
\end{figure*}

We look at temporal dynamics and then analyze the most frequent keywords and the distribution of claims across our three categories of election, COVID-19, and other. The distribution of claims into three categories is shown in Figure \ref{fig:category}b. We plotted word clouds to get an overview of the content discussed in all languages as shown in Figure~\ref{fig:wordclouds:perlanguage}.

\begin{figure*}[!htbp]
    \centering
    \subfloat[English-Hindi\label{English-Hindi-Common}]{%
      \includegraphics[width=0.49\textwidth]{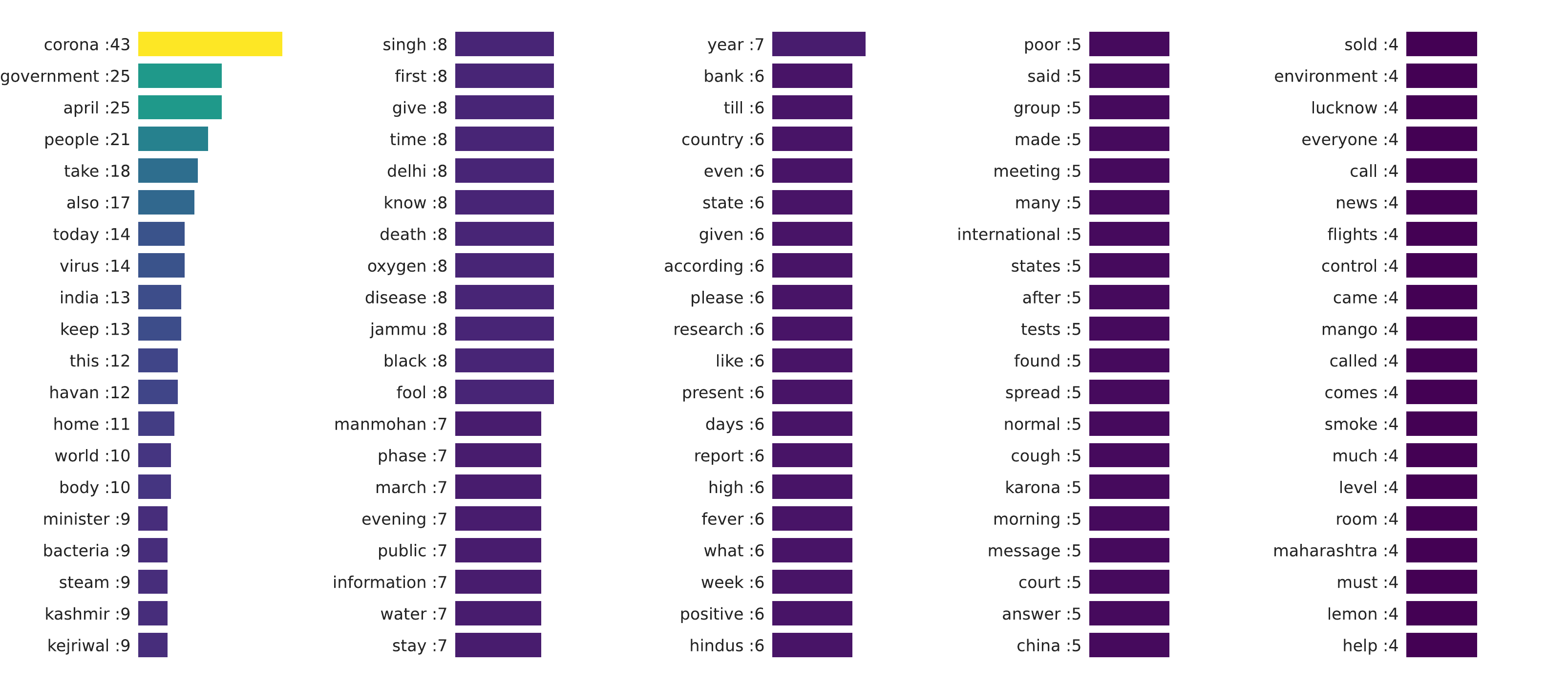}%
    }
    \subfloat[English-Telugu\label{English-Telugu-Common}]{%
      \includegraphics[width=0.49\textwidth]{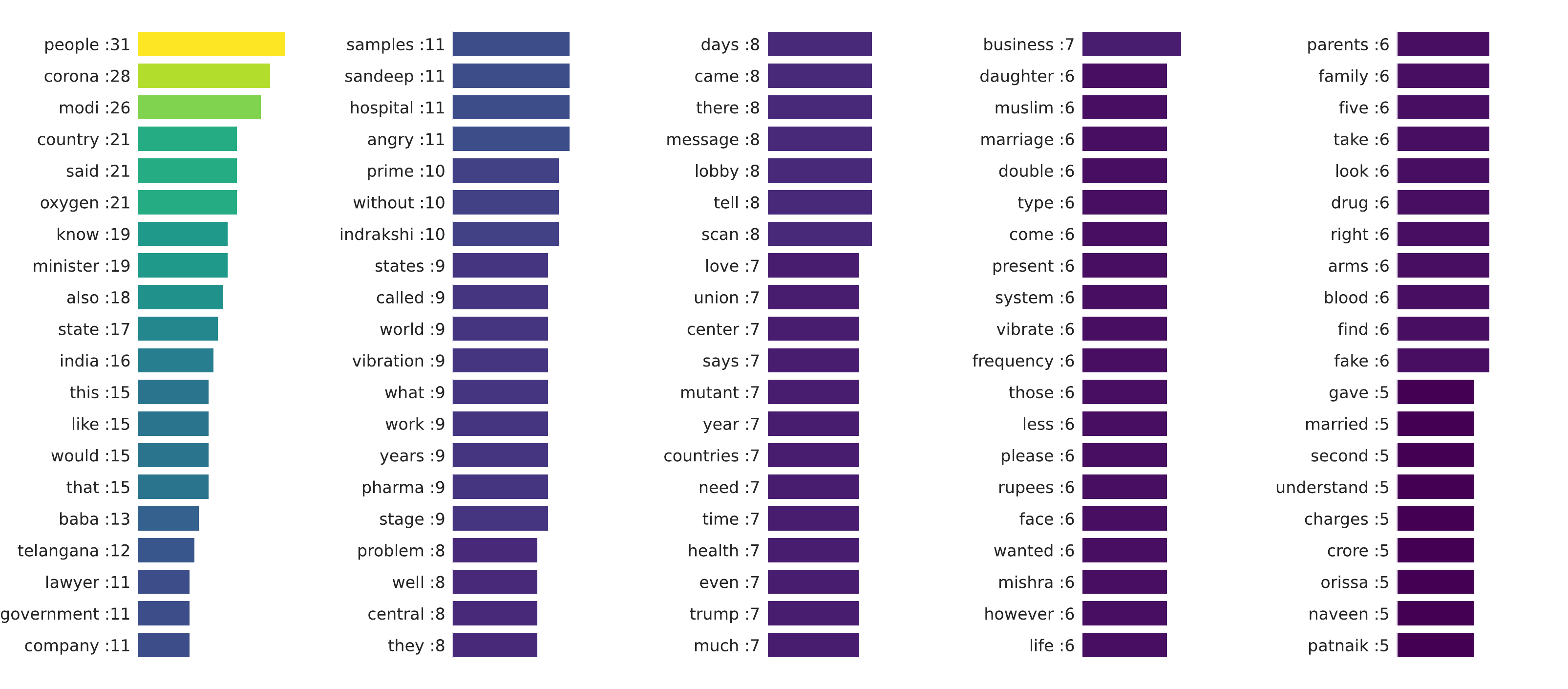}%
    }
    \hfill
    \subfloat[Telugu-Hindi\label{Telugu-Hindi-Common}]{%
      \includegraphics[width=0.49\textwidth]{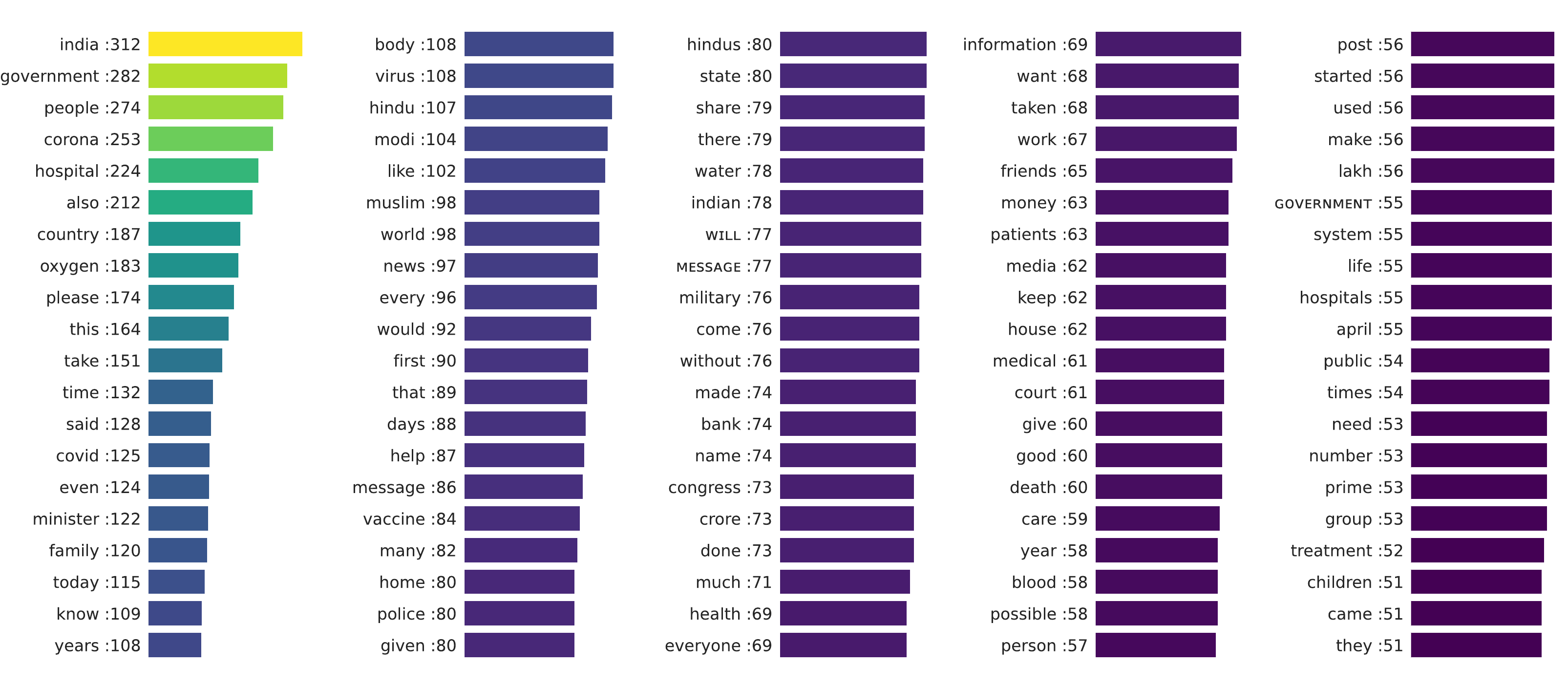}%
    }
    \subfloat[English-Hindi-Telugu\label{English-Hindi-Telugu-Common}]{%
      \includegraphics[width=0.49\textwidth]{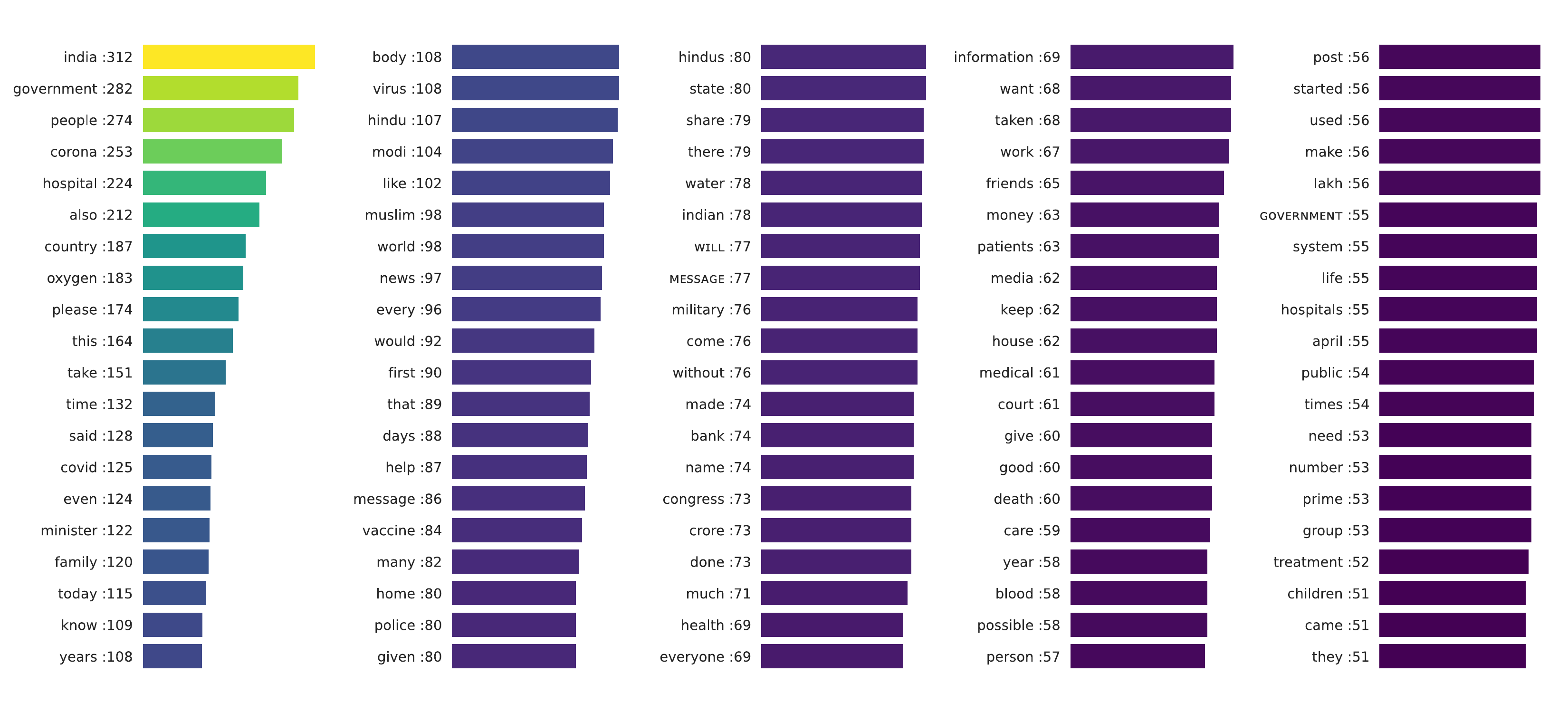}%
    }  
    \caption{Count of occurrence of shared words between different language combinations (Lighter Colors Indicate Higher Frequency)}
    \label{fig:commonword}
\end{figure*}

To answer RQ2, content across different languages, we examined frequent words using the English translations of content and found that multiple words were used in different languages, and some of the words are shared in all three languages. The overall word counts of English, Hindi, and Telugu are 5,433, 4,939, and 4,161, respectively. We computed the shared words between the two languages and found 541 words shared in English and Telugu, 500 words shared in English and Hindi, and 420  words shared in Hindi and Telegu. Overall, we have 102 words shared across all three languages. The findings show that despite cultural and language differences, many users have similar concerns in all three languages. The occurrence counts of the top 100 words shared across different language combinations are shown in Figure~\ref{fig:commonword}. The shared words from all languages discuss COVID-19, election, and other such as \textit{`mamata',`education',`smell',`ventilator',`business'}.

From the qualitative analysis of the three categories, we found more claims for COVID-19 than elections, but there is a long tail of `Other' claims. The average length of claims under the COVID-19, election, and other categories are 176, 198, and 94 words, respectively, which shows the election-related claims are longer in length. The content analysis found that multiple tipline requests have the same claim written in different ways. The distribution of these categories varies across languages as shown in Figure~\ref{fig:category}b. English has 118 COVID-19 claims, 35 election claims, and 118 other claims. For Hindi, there are 56 for COVID-19, 67 for election, and 76 for other,  while Telugu has 41 for COVID-19, 41 for election, and 28 for other. 

The content of COVID-19 is mainly about vaccines, cures, lockdowns, and postponed exams. One example cluster of claims were asking if hospitals get money to treat COVID-19 patients. For instance, \textit{``Many laymans spreading rumours saying private medical entities gets 1.5 lakhs per covid case they treat from govt  Is it true  Please write an article on this on ur website.''} 

For the election category, several topics were asked, including an increase in gas prices, one rank pension, 2G scam, money laundering scam, and religious issues. For instance, \textit{``Why don't you cover all political parties rally without mask,''} concerning the breaking of COVID-19 protocols during the election campaigns.

In the other category, the claims included international issues and asking about contact details. For instance \textit{``Bank not accepting coin then what to do'',} which is concerned about an issue but without explicit details such as bank name or location. 

\begin{figure*}[!htbp]
    \centering
    \includegraphics[width=0.94
    \textwidth]{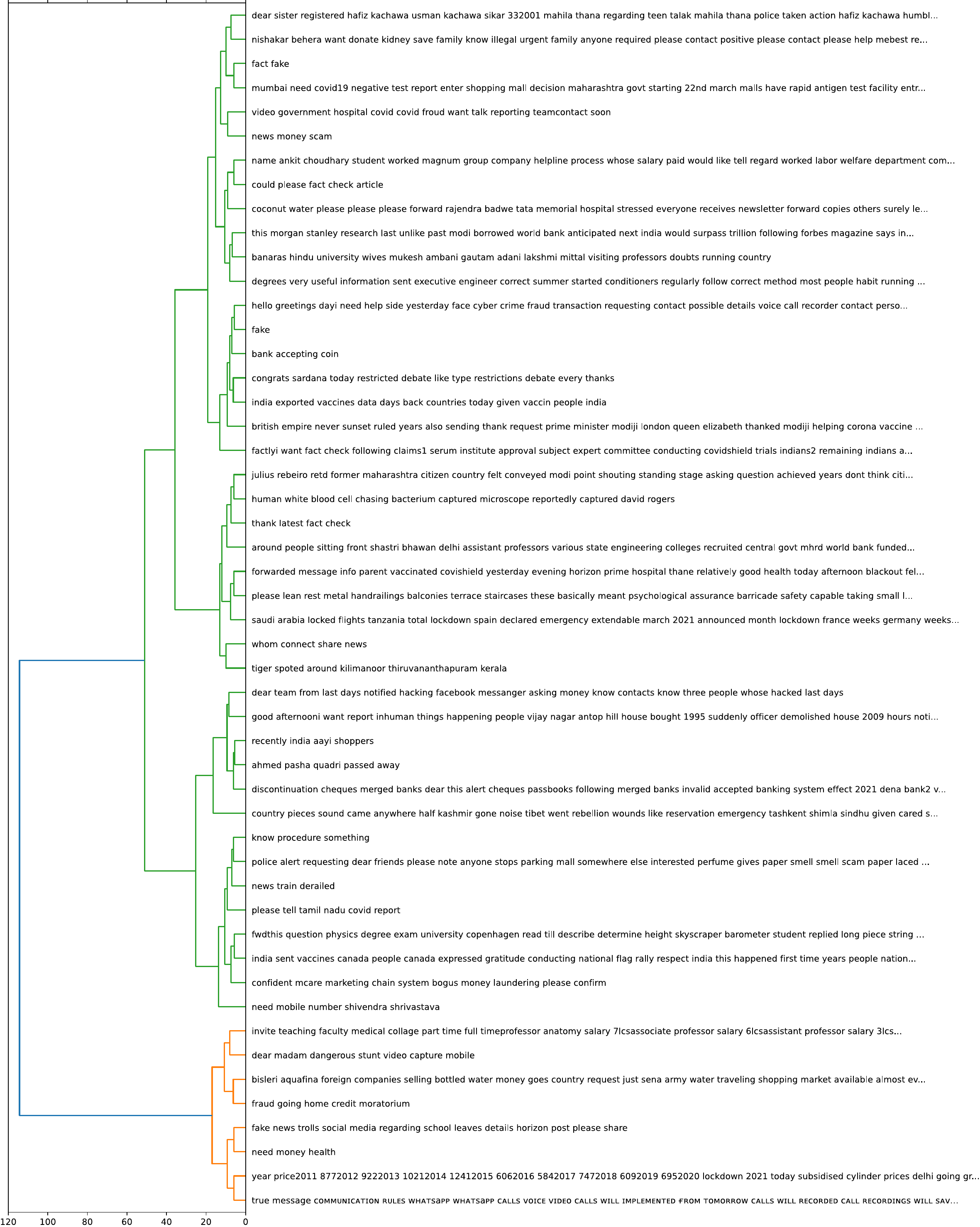}
    \caption{A Visual of Hierarchical Clustering of Tipline Claims of three languages (Hindi and Telugu claims were translated to English)}
    \label{fig:clustering}
\end{figure*}

From the semantic clustering, going from top-down, there are two main groups as shown in Figure~\ref{fig:clustering}. The first group includes a mix of topics about gas prices, mineral water prices, mobile numbers, and fraud. The second group includes election and COVID-19, such as asking for a COVID-19 report for the state of Tamil Nadu, vaccines sent to Canada, a tiger spotting, a money scam, and COVID-19 in Mumbai. Going from the bottom-up, multiple clusters formed, each discussing a different issue. Overall, there is a similarity in the categorization of issues and content from different states such as West Bengal and Tamil Nadu, which fits with our finding of similar content being requested in different languages and states. 

\begin{figure}[!htbp]
    \subfloat[Languages (Unique User Percentage Across Languages)\label{languages}]{%
      \includegraphics[width=0.5\textwidth]{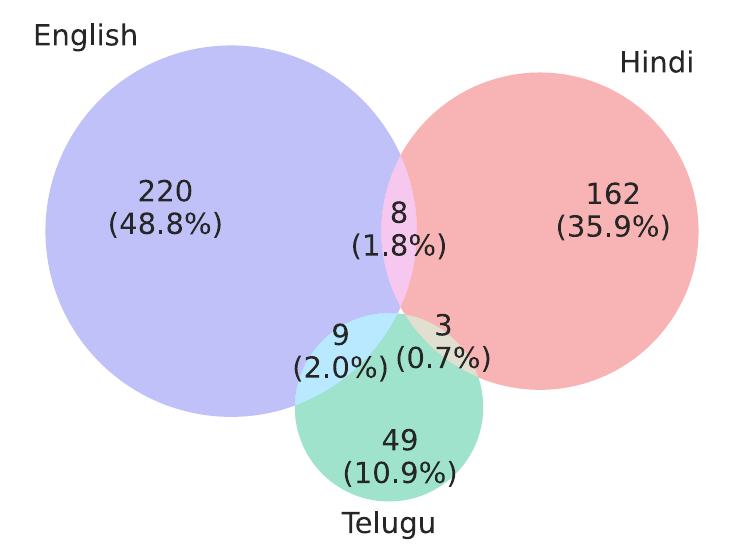}
    }
    \hfill
    \subfloat[Categories (Unique User Count Across Categories)\label{Topic of claim}]{%
      \includegraphics[width=0.5\textwidth]{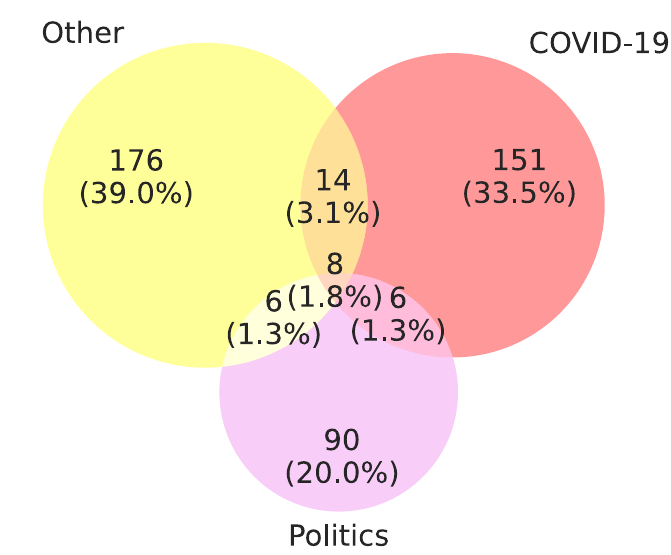} 
    }
    \caption{Repeated users based on different criteria }
    \label{fig:users}
  \end{figure}

  \begin{figure}[!htbp]
    \centering
    \includegraphics[width=.49\textwidth]{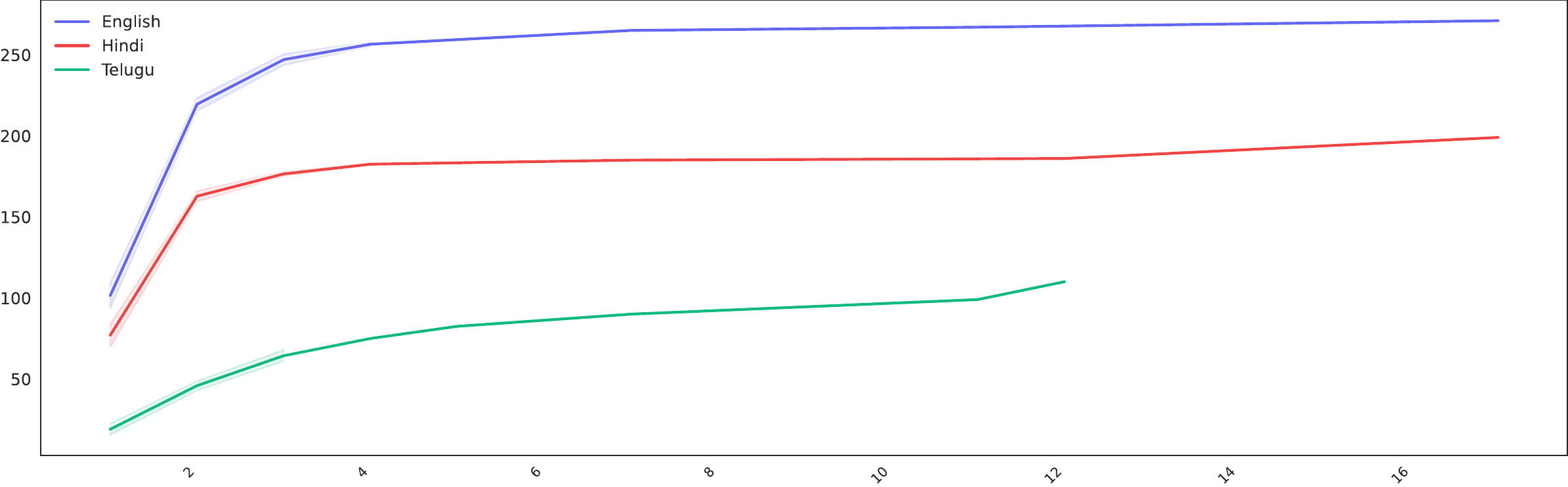}
    \caption{Cumulative request frequencies based on the number of repeat users}
    \label{fig:cummaltive}
\end{figure}

To answer RQ3, we seek to understand users' tipline requests across multiple fact-checking organizations and languages. Overall, we have 580 requests from 451 tipline users. Of these, 59 users sent multiple claims totaling 189 requests. We match the same users in two languages at a time. We found 8, 9, and 3 users in English--Hindi, English--Telugu, and Telugu--Hindi respectively, as shown in Figure~\ref{fig:users}a. We also sought to analyze the common users across different fact-checking organizations, but found there were no users who submitted claims to multiple tiplines that were fact-checked. Hence, while some users send claim requests in multiple languages, no users sent content to multiple fact-checking organizations in our study. Therefore, we infer that each fact-checking organization has an audience/userbase that is relatively independent of other fact-checking organizations. We also analyzed the users according to the category of claims. There were 14, 14, and 22 users who asked about other--election, election--COVID-19, and other--COVID-19, claims respectively. Overall, eight uses asked for all three categories. The categories of the repeated users are shown in Figure~\ref{fig:users}b. The results show that users ask for claims across languages, and they have concerns in multiple categories, especially for COVID-19. 
To analyze the relationship between users who request fact-checking multiple times and their preferred languages, we plotted a cumulative frequency graph, as shown in Figure \ref{fig:cummaltive}. The results indicate that English has the highest request volume, followed by Hindi and Telugu, with all three showing an initial sharp rise before plateauing. This suggests that a small number of users contribute significantly to the total requests, with English being the most dominant language. Also in terms of type of claims (Figure \ref{fig:category} (b)), English has the highest number of claims across all categories, followed by Hindi and Telugu, with COVID-19 and Other claims being more frequent than Politics. The results show English has diverse and frequent users who ask for fact-checking of tips.

We computed the average time spent from when a claim was first received to when the fact-check was published. This time is, on average, 2.92 days. The time spent on providing the fact-check varies: some claims require weeks to debunk, but overall around 94\% of claims are debunked within four days.


\section{Discussions}
\label{sec:d}

The 2021 Indian Assembly Elections were conducted across different states and union territories spanning multiple spoken languages. We analyzed languages, category of claims, and time to debunk. According to the societal situation, we have found more COVID-19 claims than election-related claims. 
Several claims were repeated in different regions. Besides regional issues, national issues such as COVID-19 and religious topics were heavily discussed. Figure~\ref{fig:wordclouds:perlanguage} (d),(e),(f) shows the common topics discussed between languages. Not only these, but some ongoing topics for instance Gandhi \& Nehru were also common between Hindi and Telugu. Words, for example, the world, Delhi, Jammu-Kashmir, Pakistan, and China emerged between all language pairs. This also indicates that international issues were discussed during the elections. 
Due to diversity in incoming claims requests, fact-checking needs to debunk claims on multiple topics, which is quite challenging. Resource sharing and collaboration could help in the faster debunking of claims.

\vspace{0.5ex}\noindent\textbf{Fact-checking using WhatsApp} Our contribution highlights fact-checking on WhatsApp tiplines, which allows users to send multimodal tips for fact-checking; around 37.6\% of tips were textual claims. When a fact-check is available, it is returned to the user immediately. If a fact-check is not available initially, but later becomes available users will be sent the fact-check when it becomes available. Nonetheless, we see examples where users follow up on tips they sent after a short time, indicating there may be a misunderstanding about the level of human involvement and effort in fact-checking. Spam and other out-of-scope content are also sent to tiplines, suggesting NLP pipelines to filter and prioritize content would be helpful. Tipline users should also be made better aware of the fact-checking process at work with human-led fact-checking.

In terms of verdicts of claims, the Other category is the largest in all languages as many claims sent to tiplines have no evidence or background context to provide a verdict. False and partially false are the most common ratings/verdicts. 

\vspace{0.5ex}\noindent\textbf{Multilingual Claims} WhatsApp tiplines represent a flow of claims in different languages such as English and Hindi and under-resourced languages, e.g., Telugu, which highlights that misinformation is also spread in regional languages and users request debunking of claims in their languages. Multilingual claims highlight the need for linguistic expertise because the content analysis found two major concerns about the fact-checking during WhatsApp tipline from users and fact-checkers. Users repeatedly sent several irrelevant claims that are not check-worthy~\citep{nakov2022clef} and sent reminders. For instance \textit{``what is the capital of India", ``Sir I want full news of India", ``I want to add to newsgroup"}. These claims ask for general information or news that can be found online. So, there is a need to provide knowledge to the user about the role of a fact-checking tipline on WhatsApp: i.e., that the tipline is built for fact-checking, not for news, groups, and irrelevant claims. (Or alternatively, the content on fact-checking tiplines could be expanded to include more general news.) This can be done by providing users with better guidelines, which could reduce the workload of fact-checking organizations.

\vspace{0.5ex}\noindent\textbf{Content of claims} 
We qualitatively analyzed the claims in three languages to get deeper insights and annotated them into three categories. It is clear COVID was a major topic during the 2021 assembly election with a high presence of COVID-19 claims across different languages. Overall, we found 37\% COVID-19-related tips, 25\%  elections tips, and 38\%  others. The significance is that even during crises and elections, various types of tips were requested for debunking, which varied across languages and verdicts. For instance, categorical analysis shows English has the most COVID-19 claims, i.e., 45\%. However, Hindi and Telegu have the most elections-related claims, around 38\%, which is more than the overall average. The ratings of these categories also vary across languages. Around 29\% of COVID-19 claims and 24\% of election claims are rated false or partially false. However, COVID-19 and the election also have around 12\% of claims rated true. Most of the claims in all categories are not verified due to a lack of information or being out of scope.

\vspace{0.5ex}\noindent\textbf{Significance of Low-resourced Language} 
The 2021 assembly election was held in 5 Indian states. Beyond Hindi and English, there are many regional, lower-resource languages such as Malayalam, Bengali, Assamese, Tamil, Telugu, and Urdu. However, the number of tips debunked in these regional languages is negligible, which demonstrates a need for more fact-checking work in regional languages. In addition, users may not be aware to share claims in these regional languages for fact-checking. Hence, most of the debunked claims are in English and Hindi only. 
From the analysis in the present study, it is clear that there are some content differences between larger and lower-resourced languages in terms of ratings and categories. In addition, in Telugu most of the claims are inconclusive.

\vspace{0.5ex}\noindent\textbf{Users Overlap} The overlap of users shows that users are multilingual and send tips in multiple languages. Several users sent the tips multiple times to the tipline. However, users do not send the tips to multiple fact-checking organizations. Users are explicitly using a specific fact-checking organization to debunk misleading content. Collaboration could enable tips to be better routed (e.g., between fact-checkers that focus on different languages) or identify content that another organization has already fact-checked.

\vspace{0.5ex}\noindent\textbf{Potential use of WhatsApp} In terms of using WhatsApp tiplines during future elections, our analysis shows there are misinformation claims circulating on WhatsApp. Past research has found such claims can even lead to severe harm such as mob lynching \citep{shahi2023exploratory}. An effective use of WhatsApp tiplines during the election will be beneficial in mitigating multilingual misinformation claims. While we only analyzed three languages, we found some of the data in Urdu and Bangla; so, tiplines in other regional languages, such as Marathi and Tamil, would likely be useful. This would, however, require more fact-checkers in regional languages. To avoid repetitive and noisy claims, the tipline service can provide a list of common prebunks and provide user awareness.

\vspace{0.5ex}\noindent\textbf{Practical Recommendation} From our analysis, we suggest some practical recommendations for fact-checkers and users. Our results show the importance of tools that can match claim requests and provide results such as those used in the tiplines we study. 
Adding linguistic experts who can understand and debunk claims in regional languages is required. User literacy is also needed to spread awareness of tipline-based fact-checking, which includes understanding the time required for fact-checking. In addition, users need to be made aware of what languages each organization operates in to avoid sending claims in a language that the organization does not use. 


The landscape of fact-checking on social media platforms is changing. Previously, X launched community notes \citep{prollochs2022community}, and more recently, Meta announced the end of its third-party fact-checking programs for Facebook, Instagram, and Threads, replacing them with a similar program.\footnote{\url{https://tinyurl.com/metacbf}} WhatsApp, as an encrypted platform, plays a crucial role in spreading misinformation. To enhance user awareness, WhatsApp could allow local, on-device APIs so that users could optionally choose to install apps that allowed them to flag misleading content and be alerted when seeing potentially false content. Such an approach would operate like anti-virus software using hashes for images \cite{reis2020can} or text similarity for textual claims \cite{kazemi2022research}. This would help mitigate misinformation more efficiently with the support of volunteer users.

\vspace{0.5ex}\noindent\textbf{Ethical Considerations} 
WhatsApp tiplines operate on an end-to-end encrypted platform and are entirely opt-in. Users are clearly informed about the use of their data for fact-checking and non-commercial academic research. When a tip is debunked, the fact-checking organization shares the response with user and/or publishes fact-check articles without disclosing any personal information. Throughout this research, we were extremely concerned about the privacy of tipline users and the ethical concerns. The analysis in this paper was conducted on anonymized data from which all personal information had been removed. The researchers followed strict data access, storage, and auditing procedures to ensure accountability.

\section{Conclusions \& Future Work}
\label{sec:c}

This study analyzed multilingual tipline data from WhatsApp during the 2021 Indian assembly elections. We analyzed the data from the user perspective and multilingual content in Hindi, English, and Telugu. The analysis is based on 580 filtered claims in three languages. We found that multiple users sent repeated claims, and similar claims were discussed across different languages. The timelines of incoming claims was aligned with the election and cases of COVID-19 in India. 
We have found some recommendations to deal with problems with incoming tips and provide a rating. Overall, there is a need for fact-checking organizations to collaborate across different regions on multilingual content because, despite having regional issues, national issues are present during the election and crises such as COVID-19. 
While annotating the tips into different categories, we found that around 38\% of claims were annotated as other, which shows WhatsApp users have concerns about other topics even during elections and crises.

As a limitation, our final sampled data is small in size, and we could increase the data by examining claims that were not fact-checked, expanding the time window, or involving more fact-checking organizations. We have used language translation, which also brings some errors to the text. To compare texts across different languages, we used the Google Translate API, which made some mistakes in the translation and could not translate some very long texts, which we manually fixed. For instance, a Hindi Latin claim \textit{Kya Maharashtra lockdown huwa hai} was wrongly translated as \textit{Is Maharashtra Lokhadun?}. However, we manually fixed this translation to \textit{Is there a lockdown in Maharashtra?}. Despite our manual review, it is possible some errors may remain in the data. 
We only had access to anonymous data (no phone numbers and no names); so, it is not possible to conduct any follow up interviews or relate the data to other datasets such as opinion polls.

In the future, we hope to work with fact-checking organizations to analyze tipline data during different elections or crises and compare those results to these. Future work should also include more low-resource, regional languages. Cooperation with fact-checkers can also extend the analysis to other data formats, e.g., images, or across nations. 
Finally, research should investigate how various factors influence the time fact-checkers need to debunk a claim, such as the claim's complexity, topic, and language. Another direction could be to investigate how users engage with debunked claims and help in mitigating misinformation claims.






\section*{Declarations}


\begin{itemize}
\item \textbf{Conflict of interest/Competing interests} 
Meedan is a technology non-profit that develops open-source software that fact-checking organizations use to operate misinformation tiplines on WhatsApp and other platforms. The research team at Meedan operates independently: the research questions, approach, and methods used in this article were decided by the authors alone.

\item  \textbf{Ethics approval and consent to participate} 
Throughout this research, we were extremely concerned about the privacy of tipline users and the
ethical concerns that come with such studies. All WhatsApp tipline claims in our datasets were
pseudoanonymous and we did not have access to any personal information (e.g., phone numbers). Our experiments were
done at the macro level, and we followed strict data access, storage, and auditing procedures to ensure
accountability.

\item \textbf{Consent for publication}
Both authors have approved the content of the manuscript. All authors have given explicit consent to publish this manuscript. 
\end{itemize}


\printcredits

\bibliographystyle{cas-model2-names}

\bibliography{fact_2024}


\bio{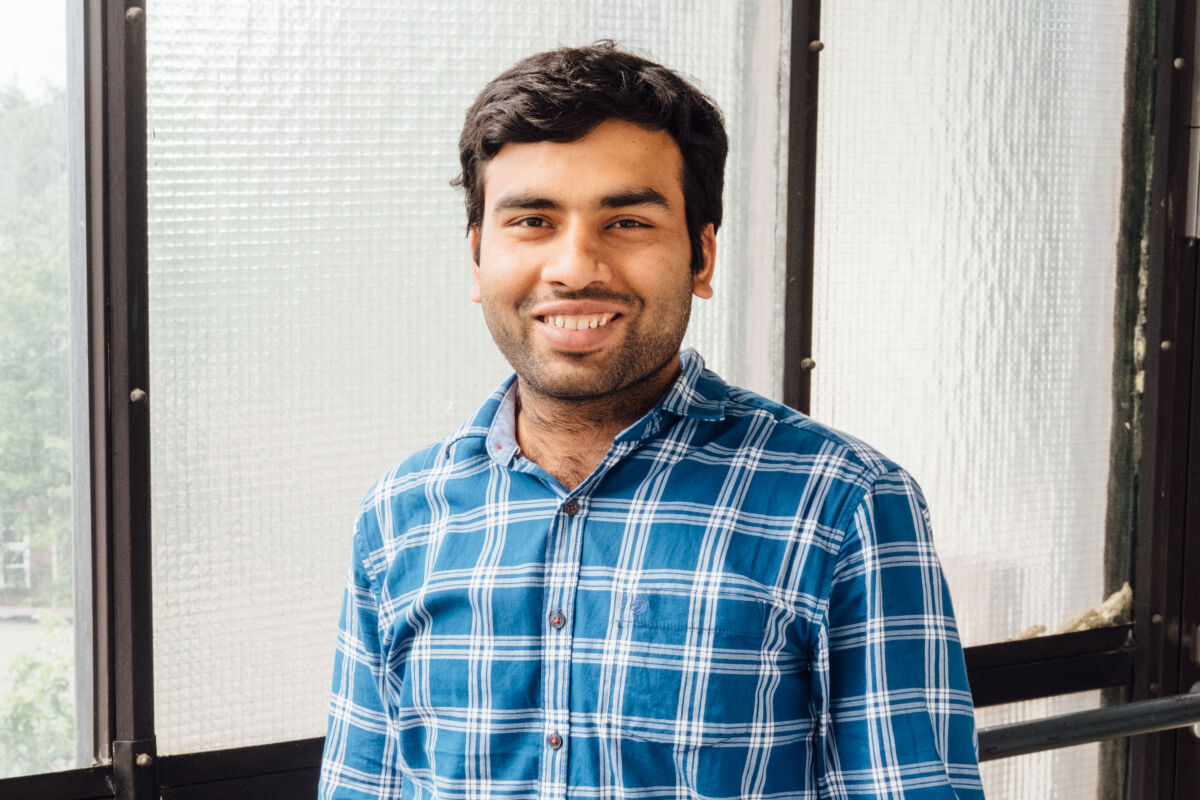}
Gautam Kishore Shahi is a PhD student at the University of Duisburg-Essen, Germany. His research interests are online harmful content, fact-checking, and Social Media Analytics. He has a background in computer science engineering with a focus on data science. His PhD is focused on the diffusion of harmful content on social media. Gautam received a master’s degree from the University of Trento, Italy, and a bachelor’s Degree from BIT Sindri, India.
\endbio

\bio{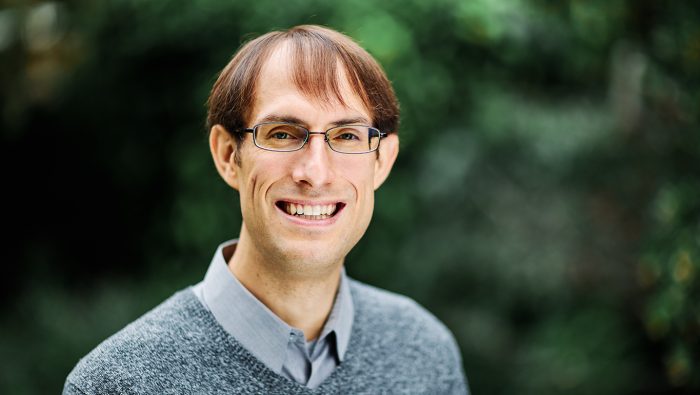}
Dr Scott A.\ Hale is an Associate Professor and Senior Research Fellow at the Oxford Internet Institute, University of Oxford, and Director of Research at Meedan. He develops and applies techniques from computer science to research questions in the social sciences. His research seeks to see more equitable access to quality information and investigates the roles of bilingual Internet users, LLM alignment, collective action and mobilization, hate speech, and misinformation.
\endbio

\end{document}